\documentclass[acmsmall]{acmart}

\AtBeginDocument{%
  \providecommand\BibTeX{{r
    \normalfont B\kern-0.5em{\scshape i\kern-0.25em b}\kern-0.8em\TeX}}}

\acmJournal{TOCHI}
\acmVolume{X}
\acmNumber{X}
\acmArticle{X}

\usepackage[final]{pdfpages}

\usepackage{xcolor}
\usepackage{wrapfig}

\definecolor{Purple}{RGB}{75,0,130}
\definecolor{LightBlue}{RGB}{135,206,250}
\definecolor{Blueish}{RGB}{30,144,255}

\newcommand{\alissa}[1]{}
\newcommand{\petr}[1]{}
\usepackage{subcaption}
\usepackage{cleveref}
\usepackage[inline]{enumitem}
\usepackage[utf8]{inputenc}
\usepackage{multirow}

\usepackage{graphicx}
\usepackage{wrapfig}

\begin{document}

\title{Designing for emotion regulation interventions: an agenda for HCI theory and research}


\author{Petr Slovak (*)}
\affiliation{\institution{King's College London}}
\email{petr.slovak@kcl.ac.uk}

\author{Alissa N. Antle (*)}
\affiliation{\institution{Simon Fraser University}}
\email{alissa_antle@sfu.ca}

\author{Nikki Theofanopoulou}
\affiliation{\institution{King's College London}}
\email{nikki.theofanopoulou@kcl.ac.uk}

\author{Claudia Daudén Roquet}
\affiliation{\institution{King's College London}}
\email{claudia.dauden_roquet@kcl.ac.uk}

\author{James J Gross}
 \affiliation{\institution{Stanford University}}
 \email{gross@stanford.edu}
 
\author{Katherine Isbister}
 \affiliation{\institution{UC Santa Cruz}}
 \email{katherine.isbister@ucsc.edu.com}

\renewcommand{\shortauthors}{}

\newcommand{\qqq}[1]{\textrm{``#1''}}

\begin{abstract}
\end{abstract}

\ccsdesc[500]{Human-centered computing~HCI theory, concepts and models}
\keywords{emotion regulation, mental health, technology-enabled intervention, review}



%
%

\begin{abstract}
There is a growing interest in HCI to envision, design, and evaluate technology-enabled interventions that support users' emotion regulation. This interest stems in part from increased recognition that the ability to regulate emotions is critical to mental health, and that a lack of effective emotion regulation is a transdiagnostic factor for mental illness. However, the potential to combine innovative HCI designs with the theoretical grounding and state-of-art interventions from psychology has yet to be fully realised. In this paper, we synthesise HCI work on emotion regulation interventions and propose a three-part framework to guide technology designers in making: (i) theory-informed decisions about intervention targets; (ii) strategic decisions regarding the technology-enabled intervention mechanisms to be included in the system; and (iii) practical decisions around previous implementations of the selected intervention components. We show how this framework can both systematise HCI work to date and suggest a research agenda for future work. 

\end{abstract}

\maketitle

\section{Introduction}

Over the past decade, the HCI literature has seen a growing interest in technology-enabled interventions to support emotion regulation \cite{Crossman2018,Slovak2018,Antle2015,Antle2018,Antle2019,Miri2020}.  
	This rapid growth is fuelled by several factors, including an increased interest in the possibilities of technology-support in the context of mental health \cite{Sanches2019,Torous2018d,Mohr2017a,Hollis2018}, the proliferation of low-cost wearable sensors enabling potentially widely deployable bio-feedback systems (e.g., \cite{Peake2018,Antle2018}), and the associated focus on combining affective computing and personal informatics agendas  (e.g., emotion awareness systems \cite{Hollis2018a,Bakker2018}) -- see also \cite{Sanches2019, Epstein2020} for two recent reviews. 
	An even more dramatic explosion of interest in emotion regulation is happening outside of HCI. Research concerning emotion regulation is one of the fastest growing areas in psychology \cite{Gross2015PsychologicalProspects}, and emotion regulation is now widely understood to be a key protective factor supporting personal well-being \cite{Aldao2010,Compas2017}. Indeed, in recent years, emotion regulation has begun to be seen as a likely trans-diagnostic intervention across a range of mental health disorders \cite{Sloan2017,Sakiris2019,Sloan2017a,Cludius2020,Musiat2014}.


Much of the existing HCI work has focussed on exploring the potential design space and establishing the promise of digital emotion regulation support across a range of contexts and populations. 
So far, however, the potential to combine the innovative HCI intervention approaches and state-of-art psychological interventions hasn't yet been fully realised (cf., also \cite{Bettis2021}). 
We see this as a part of a natural progression common to many emerging HCI fields (e.g., behavioural change systems \cite{Hekler2013}, affective computing \cite{Calvo2015}, or personal informatics \cite{Epstein2020}) whereby the HCI research community progressively moves from technology-led explorations and interaction-focused studies of proof of concepts, towards the development of robust systems and evaluation of interventions measuring large-scale, real-world impacts. This more mature stage of HCI research requires establishing supports for an interdisciplinary community in order for such systems to be designed, evaluated, and deployed with efficacy. Our aim in writing this paper is to put forward one of these supports.

%

%

\begin{figure}
    \centering
    \includegraphics[width=0.98\textwidth]{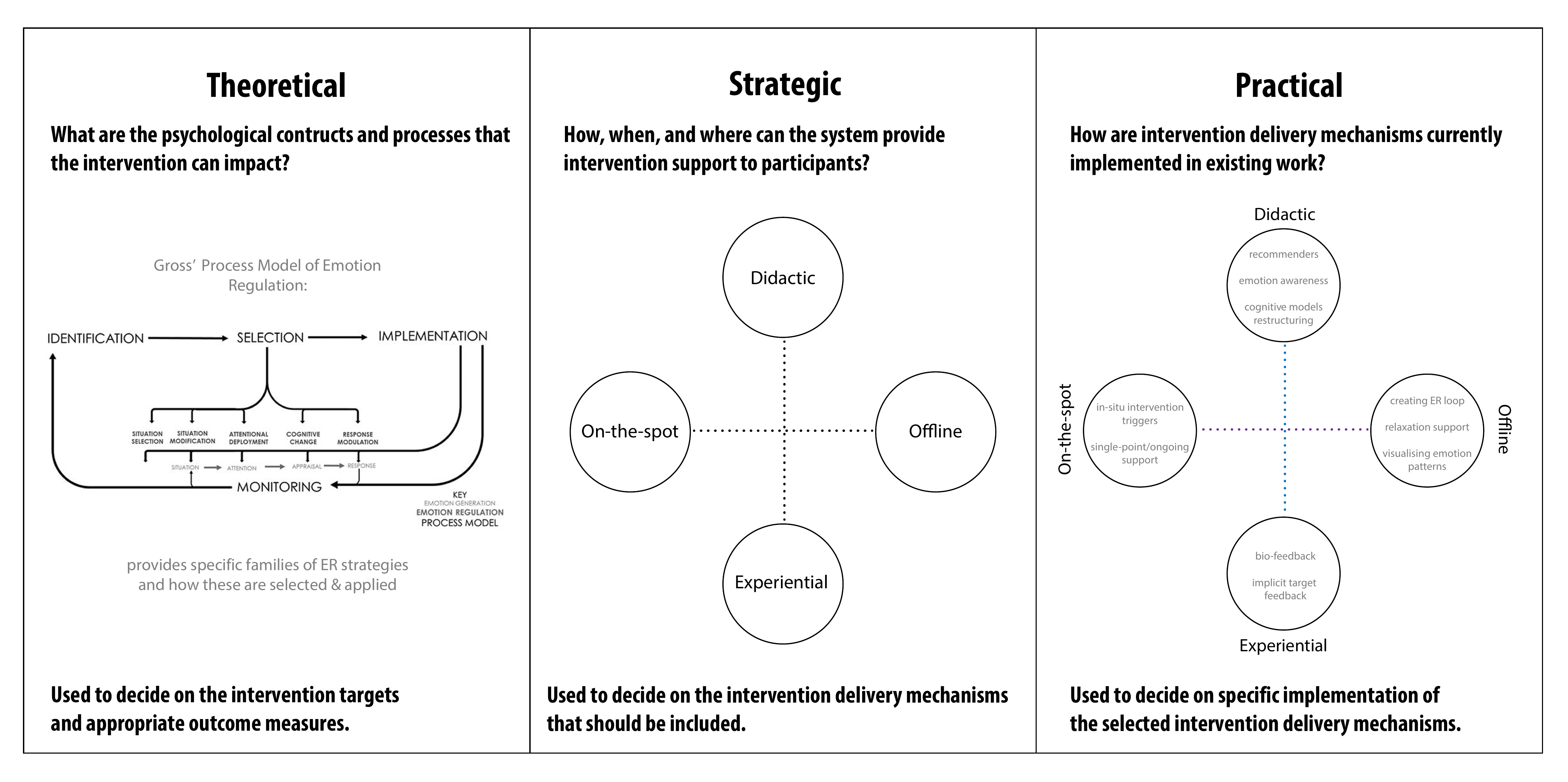}
    \caption{Overview of the three-part framework, including key questions and the role components can play in the intervention design process.}
    \label{fig:framework}
\end{figure}

In this paper, we aim to take stock of the HCI work on \textit{emotion regulation} so far, synthesise and make accessible the breadth of existing work on emotion regulation in psychology, as well as suggest promising directions for future research on technology-enabled emotion regulation interventions that the HCI community can engage in as the field matures. 

While we focus specifically on \textit{technology-enabled emotion regulation intervention systems only}, this review and agenda-setting work should be seen within the context of other discussions in the field: these include, for example, a recent review of affective health and HCI \cite{Sanches2019}, an analysis of how digital technologies that are \textit{not specifically} designed as interventions are used for emotion regulation \cite{Wadley2020}, as well as broader analyses of HCI in the context of well-being \cite{Blandford2019}, behaviour change \cite{Epstein2020}, or socio-emotional learning \cite{Slovak2017}. It also complements a series of recent workshops on emotions and HCI at UbiComp \cite{Tag2021} and at CHI \cite{Wadley2022}. 

In the rest of the paper we encapsulate the synthesis of current---and potential future---HCI work on emotion regulation interventions in a three-part framework (see Figure 1), aiming to help technology designers separate out decisions on three inter-related aspects of intervention development: 
\begin{enumerate} 
	\item \textit{Theory-informed} decisions about \textit{intervention targets} from psychology: i.e.,  `what' are the emotion regulation intervention goals the designed system could aim to impact. 
	\item Key \textit{strategic} decisions about the \textit{intervention delivery mechanisms} that should be included in their system: i.e., `how' and `when/where' can interaction design play a role in the intervention process. Specifically, we highlight the differences in designing for different approaches to \textit{how} emotion regulation skills can be developed (experiential vs didactic); and \textit{when/where} such support is offered (on-the-spot vs offline). 
	\item \textit{Practical} decisions around the commonly used \textit{design components} that represent specific implementations of selected intervention delivery mechanisms (experiential,  didactic, offline, on-the-spot), e.g., drawing on or extending the current \textit{best design practice} as identified across HCI work so far. 
\end{enumerate}

Each aspect of the framework brings a different contribution to the HCI literature: 
The \textit{theory aspect} aims to reframe and bring together important theories needed to underpin the design of technology-enabled interventions from across psychology, in ways that are easily accessible to HCI researchers. 
The \textit{strategic aspect} introduces four complementary delivery mechanism approaches, grounded in psychology. This aims to orient designers to consider decisions around the role that technology should play in the intervention system---i.e., how, when, and where technology components could affect users, rather than immediately focusing on the specific technologies that can be used to deliver psychological impact. 
%
Finally, the \textit{practical aspect} provides a snapshot of the currently common intervention design components which synthesises how support for each delivery mechanism is implemented in existing work (both technologically and methodologically), and thus also allows us to identify gaps and potential new research areas in this exciting space. 

\section*{Executive summary}

Given the length of this paper---resulting from synthesis across a range of domains and an in-depth review of existing HCI work---we start with an `executive summary' of our full argument, while referring to the respective sections for details. Our aim is that the interested reader can pick and choose where more detail is needed (and read the corresponding section) and where a simple overview is sufficient (and thus rely on the executive summary only for that part of the argument). 
In particular, 
\begin{itemize}[topsep=4pt]
\item[-] \textbf{Section 2} provides an overview of the psychological models underpinning the theoretical part of the framework. 
\item[-] \textbf{Section 3} builds on these models as well as the learning sciences literature to derive the strategy part of the framework, i.e., the `hows' (experiential vs didactic) and the `whens/wheres' (on-the-spot vs offline) of possible delivery mechanisms. 
\item[-] \textbf{Section 4} then outlines a scoping review of existing HCI work on emotion regulation to date, to ground the practical component of the framework. We identify and categorise all the interaction designs used for each individual delivery mechanism and then analyse how these individual design components are commonly combined into `interventions' in existing HCI work. 
\item[-] \textbf{Section 5} finally highlights the gaps we identify through the scoping review and showcases how an HCI researcher could pick different aspects of the three components of the framework (theoretical, strategic, practical) to guide their research at different levels, whether that is support for a specific intervention design, an individual research programme, or broader field building.  
\end{itemize}

\subsection*{Theory part summary (Section 2)} 
Psychology literature provides a useful set of \textit{theory-driven intervention targets}: i.e., well defined conceptual processes that technology-enabled interventions could aim to affect. We highlight Gross' Process Model of Emotion Regulation \cite{Gross2015}, which allows designers to think systematically about \textit{how emotions are generated from `person-situation' interactions}, \textit{the approaches users might take to impact their emotions}, and provides a \textit{set of clear intervention points} to target. 

In particular, the Process Model differentiates the intertwined processes of emotion generation and emotion regulation. Emotion generation is a result of: people \textit{encountering}, consciously or subconsciously \textit{attending to}, \textit{evaluating}, and \textit{responding to} internal (e.g., a memory/thought) or external (e.g., workplace disagreement) \textit{situations}. In contrast, emotion regulation occurs when people evaluate their current emotional state relative to their goals and decide to modify them (or not). Specifically:

\begin{itemize}
\item The process of emotion regulation is seen as proceeding in \textit{4 stages}: \textit{identification} (of need to emotion regulate), \textit{strategy selection} (from the 5 families below), \textit{implementation} (of a particular instance of a strategy within the current context), and \textit{monitoring}  (whether to continue, stop, or change ER strategy).  	

\item The 5 families of specific ER strategies then synthesise the common approaches that people can draw on to up-/down- regulate an emotion at a given moment, each corresponding to an appropriate step in the emotion generation model (\textit{situation selection, situation modification, attentional deployment, cognitive change, response modulation}).
\end{itemize}
Individual technology-enabled intervention components can thus focus on either providing support for any specific strategy (e.g., cognitive change) and/or the overarching emotion regulation stages (e.g., supporting emotional awareness to enhance the identification/monitoring stage).  
We note, however, that the evidence from existing clinical programmes suggests interventions are most likely to be effective if they support a \textit{range of emotion-regulatory processes} (rather than a single set of isolated interactions or strategies), suggesting the need to combine multiple (technology-enabled) intervention components into a broader intervention package if real-world impact on users' ER capabilities is a key concern.%

Another important implication from the psychological literature is that learning to regulate emotions is best thought of as supporting the development of an \textit{embodied skill} (rather than, say, an ‘information delivery’ problem). From the interaction design perspective this means that the ability of the system to either \textit{help elicit target emotions} through an interactive experience, or \textit{to support users during  everyday emotional situations} is often crucial to help users develop their emotion regulation capabilities fully, and to successfully transfer the new knowledge to everyday settings. 



\subsection*{Strategic part summary (Section 3)} 
It is unlikely that---as HCI researchers---we would aim to contribute to the fundamental understanding of intervention targets or psychology of emotion. 
Instead, our main contribution will likely lie in envisioning new ways in which these theory-driven intervention targets can be scaffolded through innovative intervention delivery mechanisms.  

The strategic component of the proposed framework thus aims to help guide designers in deciding \textit{how, when, and where the intervention support} could be most effectively offered through technology. 
%
In particular, we base the strategy components on learning sciences literature to encourage HCI designers to think about delivery mechanisms in the context of their role in the intervention process (i.e., when, where, and how the design \textit{scaffolds participants' learning}), rather than defining the intervention approaches in the context of specific technologies or interface-level characteristics: 

\begin{itemize}
    \item[-]  The \textit{hows} part of the framework distinguishes \textit{the type of skills acquisition approach} an intervention component supports: it highlights the difference between \textit{didactic learning} (such as the information delivery based approach commonly used in school and delivered e.g., through reminders)  and  \textit{experiential learning} (such as the trial\&feedback approach commonly used in mindfulness and delivered, for example, through biofeedback). 
    \item[-]  The \textit{whens/wheres} part of the framework then helps the designers think about \textit{timing and contexts in which intervention support is provided} through technology: distinguishing between  \textit{`offline' training} (such as intervention support in a therapist office or an online module); and \textit{`on-the-spot' learning} (i.e., intervention support provided during everyday emotional situations in-situ).
\end{itemize}

We note that any resulting intervention could be composed of multiple complementary delivery mechanism components: for example, a game-based bio-feedback intervention (such as \cite{Knox2011}) can be seen as utilising an `offline' delivery mechanism through the game (which is used to generate emotions outside of everyday situations), and an `experiential' mechanism (by involving the biofeedback loop), as well as perhaps a `didactic' component (if, for example, the participants first engage in psychoeducation before the game starts). 

As argued in more depth in Section 5, we propose that this conceptual structure can enable a modular thinking about intervention design (including a 'dictionary' of common design components outlined in the next section) which provides a framework for the community to develop an empirically grounded understanding of `what works' for each type of delivery mechanism and promotes cross-project inspiration across HCI and psychology research teams. 
%

%


\subsection*{Practical part summary (Section 4)}
The role of the practical part is to provide the designers with a synthesis of the existing work on emotion regulation technologies in HCI to date to guide practical decisions on possible implementations of selected delivery mechanisms, as well as to help identify areas where current solutions need improvement or are missing. In other words, the practical component comprises:
\begin{enumerate}
    \item A synthesis of the commonly used \textit{intervention design components} used to support each type of delivery mechanism within existing HCI work (i.e., a ‘dictionary’ of techniques used so far for experiential, didactic, offline, on-the-spot support).
    \item An overview of how these design components \textit{are currently combined into interventions} (i.e., which combinations of delivery mechanisms and their implementations already have been explored in the literature). 
    \item An overview of \textit{the range of the ER targets} within the current work (i.e., which of the 4 ER stages / 5 families of ER strategies have been supported so far).  

\end{enumerate}

In contrast to the theory and strategic parts of the framework, the practical part will always be just a `snapshot’ in time and will require ongoing updates to incorporate newly published work. The descriptions that follow are based on a scoping review including papers published in HCI literature between January 1 2009 and December 8 2021 (cf., \cite{Munn2018_ScopingReview}), as primarily indexed across ACM digital library and IEEE, and complemented with an additional Google Scholar search.\\
{
\small 
\textit{NOTE - The authors (i) will update the review for camera ready version; as well as  (ii) maintain an up-to-date database of papers for at least 5 years after the publication on OSF (as part of Slovak’s FLF fellowship).}
}

\subsubsection*{Design components}\quad  Overall, the analysis shows that the existing work in HCI is, so far, relatively coherent with only a small number of interaction design approaches supporting either the type of skill acquisition approach (didactic vs experiential; i.e., 'hows') or the intervention delivery (on-the-spot vs offline, i.e., 'whens/wheres').

\begin{itemize}
\item For the \textit{`hows'}, \underline{experiential} design components focus predominantly on \textit{bio-feedback} or \textit{implicit target feedback} (i.e., nudging the users (sub-)consciously toward particular physiological states through haptic interaction). In contrast, \underline{didactic} intervention components draw on \textit{reminder-recommender} systems (i.e., suggesting specific ER strategies to users), facilitate \textit{emotion awareness} (e.g., prompting users to think about how they feel and/or felt), and rely on traditional \textit{cognitive models restructuring} (often based on established psychoeducation approaches from existing therapies with limited additional technology support). \smallskip

\item For the \textit{`whens/wheres'}, the \underline{offline} design components aim to use interactivity to create \textit{a bespoke emotion regulation loop} (e.g., using a video game to elicit feelings of stress while, at the same time, providing other support such as a biofeedback loop to down-regulate those feelings); to facilitate \textit{relaxation training support} (similar to above but without the emotion generation component); and to scaffold users' reflection by \textit{visualising patterns of emotions over time} (e.g., by showing a timeline of physical locations and emotions felt at each). The \underline{on-the-spot} components then draw on a range of \textit{intervention triggers} (to start and/or support users' in-the-moment emotion regulation process), and can offer either \textit{single-point} (e.g., reminder) or \textit{ongoing} support (e.g., haptic feedback throughout a stressful experience).
\end{itemize}

\subsubsection*{How are these combined into interventions?}\quad
When looking at how these intervention design components are combined into final systems the existing work clearly clusters into three main research areas, as outlined below. 

\begin{itemize}[topsep=2pt]
\item[-] \textit{implicit on-the-spot support} (e.g., providing a target breathing rate through haptic input; experiential + on-the-spot); 
\item[-]  \textit{bio-feedback games} (e.g,. embedding a biofeedback component into stressful gaming as a visual overlay; experiential + offline); and 
\item[-] \textit{reminder-recommenders \& awareness systems} (e.g., push notifications suggesting ER strategies; didactic + on-the-spot). 
\end{itemize}
Interestingly, these approaches are mostly independent from each other as they draw on different background literature, intervention goals, and technological components.
This clustering also neatly follows the associated combination of used delivery mechanism approaches as well as consistent combinations of components within each cluster (i.e., design components used in one cluster, e.g., implicit feedback or recommenders, are rarely utilised in other clusters).

Overall, these observations suggest that the work so far has been fragmented along technology-specific sub-domains (e.g., biofeedback, reminders) rather than the focus on ER intervention aspects. This suggests an under-utilised opportunity to combine and build on existing knowledge.
%
%
Additionally---as is traditional in HCI---the contribution of many of the papers to date is aimed at early stage exploration of innovative design spaces, rather than large scale deployments or experimental/randomised controlled trials that would be required to test the effects of designed systems on changes in emotion regulation skills (cf., \cite{Sanches2019}. As such, while we are starting to develop an understanding of which interaction design mechanisms are engaging for the users, the field still lacks substantive data on long-term efficacy of the individual techniques and their impact on users' everyday emotion regulation/well-being. 

%

\subsubsection*{Common ER targets}\quad There is a similar lack of diversity of focus when looking at the ER intervention targets (i.e., the `whats'): the majority of work across implicit on-the-spot support and bio-feedback games areas targets the \textit{response modulation} family of strategies. Any additional ER strategies are introduced through simple psychoeducation components (e.g., based on traditional CBT), mostly without added technology support. 
The reminder-recommender and awareness systems mostly do not teach or support specific emotion regulation strategies, but instead focus on supporting \textit{identification} and/or \textit{situation selection} as a specific strategy, or simply point to a range of possible ER strategies without further scaffolding. 

Importantly, the majority of papers across our whole sample (31/36) assume that the developed systems have to provide an on-going support for the intervention to be effective; and thus that the effects would disappear if the system (such as a wearable feedback system) were removed. In other words, most of the existing HCI systems approach emotion regulation as something that requires on-going technological scaffolding, rather than a skill that the 
user could develop (with a temporary technology-enabled intervention). 



\subsection*{Research agenda and envisioned framework use (Section 5)}

The proposed framework is our way of communicating both the understanding of what has been done so far (cf., Section 4) \textit{and} a way of helping shape future research at the intersection of the HCI and psychology fields (Section 5). 
We show how the three parts of the intervention development framework (theory, strategic, practical) can be applied---separately or together---as \textit{design lenses} to identify gaps in existing work, inform new projects, and support ongoing synthesis of work across the range of disciplines and possible research threads. 
Section 5 describes three such applications: 

\subsubsection*{ER targets gaps}\quad First, we focus on the gaps in technological support for possible ER targets and psychological impact of our systems (theory \& practical). We identify 3 key themes: the \textit{lack of transfer support} (i.e., only a few systems so far designed 
to empower users to develop emotion regulation competence); \textit{uneven support for ER strategies} (i.e., predominant focus on supporting response modulation, while strategies in support of other ER strategies remain under-researched); and \textit{lack of efficacy data} (i.e., we have only very limited understanding of the actual psychological impact of the HCI systems designed so far). 

\subsubsection*{Intervention mechanism gaps}\quad Second, we foreground the gaps in our knowledge of effective support for each intervention delivery mechanism (strategy \& practical). For each of the four delivery mechanisms (didactic, experiential, on-the-spot, offline) we suggest 2-3 research questions that have yet to be addressed in current work but are---in our minds---crucial for fully developing the potential of technology-enabled intervention support (see sections 5.2.1 -- 5.2.4). For example, the experiential mechanism questions point to the lack of research on supporting \textit{performative aspects of well-known adaptive ER strategies (e.g., cognitive change, attentional deployment)}; and the limited research on \textit{how technology could directly scaffold and/or guide users through the emotion regulation experience} in-situ.

\subsubsection*{Connecting interdisciplinary communities}\quad Finally, we argue that---within such an inherently interdisciplinary space---it is important to support practitioners at varying scales of granularity, recognising the importance of research engagement at all levels of interest, resources, and research goals (e.g., from in-lab technology innovation that aims to showcase a proof-of-concept for a new interaction mechanism; to real-world large scale intervention deployments that aim to understand robust intervention efficacy across user contexts). We provide one possible outline of structuring such progression from early HCI innovation work to psychological application, and point to examples from other domains where similar approaches have been successfully employed.

\vfill \pagebreak

\section{Theory component -- emotion regulation models (the `whats')}
\label{sec:step1-all}
%
Emotion regulation is broadly defined as "\textit{the set of strategies that individuals may use to increase, maintain or decrease their affective experience, including the feelings, behaviours or physiological responses that make up a given emotion \cite{ER-handbook-Gross-14}}". The purpose of this section is not to provide a full review of all that is known about emotion regulation, as that is simply infeasible given the incredible growth of the literature of the last two decades; see  \cite{Sheppes2015,England-Mason2020,Sloan2017b,Southam-Gerow2002,Kneeland2016,Compas2017} for recent reviews. Instead, we aim to synthesise key aspects of the most influential\footnote{The research on this model was reviewed in major psychology journals---for example Psychological Inquiry \cite{Gross2015} and Annual Reviews of Psychology \cite{Sheppes2015}---and these two review papers alone have received more than 3500 citations by the end of 2021.} emotion regulation model---Gross' Process Model of Emotion regulation \cite{Gross1998a}---especially insofar as these can be generative for the design of emotion regulation technologies and help prepare ground for the psychology-informed mapping of existing HCI systems in the next section. 

%

\subsection{Psychological models of emotion regulation} 
\label{sec:emotion-def}
%

\begin{figure}
    \centering
    \includegraphics[width=.9\textwidth]{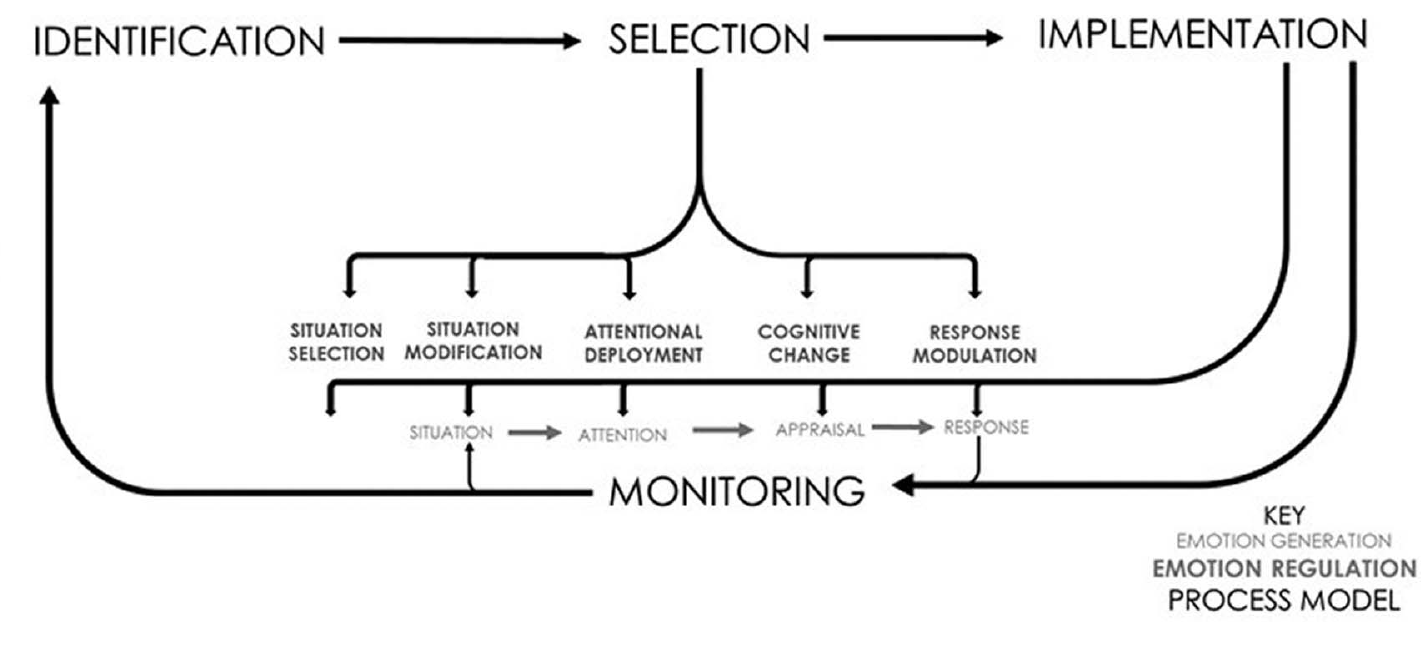}
    \caption{The Process Model of Emotion Regulation, including emotion generation and the associated families of ER strategies (adapted from \cite{McRae2020}, with permission)}
    \label{fig:ER-5-families}
\end{figure}

%
There is a broad agreement that emotions: 
\begin{enumerate}[topsep=2pt]
    \item involve \textit{person--situation transactions} that \textit{compel attention}; 
    \item have meaning to an individual \textit{in light of currently active goals}, and 
    \item give \textit{rise to coordinated yet flexible `full body' responses} (i.e., involving cognitive, behavioural, and physiological components) that may modify the ongoing person--situation transaction in crucial ways.
\end{enumerate}
%
%
Gross's model distinguishes between emotion generation and emotion regulation \cite{Wadley2020}. During \textbf{emotion generation} people encounter, attend to, appraise and respond to situations. Figure~\ref{fig:ER-5-families} shows how these features can be synthesised into a useful, if simplified model of \textit{emotion generation} (cf., \cite{Gross2015}): The respective \textit{situation} must be consciously or unconsciously attended to (\textit{attention}) and then evaluated in light of an individual's active goals (\textit{appraisal}), which then gives rise to the associated changes in subjective, behavioural, and physiological response that we perceive as emotion (\textit{response}). In contrast, \textbf{emotion regulation} occurs when people evaluate their current emotional state relative to their goals, and decide to modify them (or not). If they choose to modify their state they select a particular strategy to use, implement it with context-specific tactics, and then monitor the process and outcomes, possibly repeating one or more of these steps \cite{Gross2015}. 

This model is useful because it describes five families of ER strategies that people can use to regulate emotion based on when in the cycle they intervene in the emotion generation process (see Section~\ref{sec:families}). Within each family, strategies can be adaptive or mal-adaptive within the specific context, depending on a number of factors including effects of short and long term use (e.g. binge watching a YouTube series might be adaptive initially after a negative life event, but if continued would have longer term negative impacts, such as social isolation, reduced work productivity, and less time to develop other adaptive strategies).   
\alissa{I added a bit more detail to differentiate EG ER and tie into figures. Should be useful later in paper when we're trying to tie our analysis/findings into this model -see what you think}
We note that while the `psychologically relevant situations' mentioned in the model can often be specified by referring to features of the external environment (e.g., seeing a tree in the middle of the slope at the last moment while skiing; or watching a well-made horror movie), it is also possible for psychologically relevant `situations' to be internal (e.g., the sneaking suspicion that I'll never amount to anything).

Before moving on to outline the emotion regulation processes in more detail, it is important to highlight a key characteristic which is embedded in the definition above: arising from the `person--situation interchange', emotions are transient, and thus are often difficult to elicit without reinstating the corresponding situation in full. For example, remembering your last skiing line or a roller-coaster ride is very different to actually doing it; which is one of the reasons why skiing resorts and adventure parks are still in business. 

As we will discuss in more detail later (Section~\ref{sec:learning_sciences}), this experiential and situation-dependent nature of emotion has crucial implications for emotion regulation interventions: \textit{learning to regulate one's emotions requires having access to the emotions that are to be regulated} (in the same way as learning to ski requires being on the slopes). As emotions are hard to `re-live', the need to either elicit appropriate emotions during ER training or support in-situ application of ER strategies is one of the key difficulties intervention programs face. 
It also signposts the intertwined issues of emotion \textit{generation} and emotion \textit{regulation}, which we turn to in the next section.

\subsection{What are the basic emotion regulation processes?}
\label{sec:families}

Emotion regulation occurs when \textit{there is an activation of a goal that recruits one or more processes to influence emotion generation} (cf., \cite{Gross2015, Sheppes2015}). In what follows, we will focus on situations where the target of the regulatory goal is the person who experiences the emotions (i.e., \textit{intrinsic} emotion regulation), rather than aiming to induce change in someone else (i.e., \textit{extrinsic} emotion regulation, such as when parents scaffold a child's emotional experiences). We note that such regulation can occur for a variety of reasons: people might choose to up-regulate positive states, down-regulate negative emotions, or generally regulate their emotions for instrumental reasons (e.g., up-regulating anger during a conflict situation). Finally, an important distinction is between ER processes that are effortful and consciously enacted (i.e., \textit{explicit} ER) and ER processes that are automatic and occur outside of people's awareness (i.e., \textit {implicit} ER).
The key components of the process model of emotion regulation model have been described in detail in many other papers --- 
we refer the reader to, e.g., \cite{Gross2015PsychologicalProspects} for a full description to complement the deliberately succinct overview below and the illustration in Figure~\ref{fig:ER-5-families}.

 

\subsubsection{Overarching emotion regulation process}
In summary, emotion regulation is seen as a \textit{dynamic, ongoing valuation process} comprised of four main stages, which feed into each other -- see the outer layer in  Figure~\ref{fig:ER-5-families}: 
\begin{itemize}
	\item First is the \textit{identification} stage, which is involved in making the very general decision of whether to start regulating or not. 
	\item A decision to change an emotion state leads to the \textit{selection} stage, which decides which of currently available general 5 families of emotion regulation strategies is to be used: situation selection, situation modification, attentional deployment, cognitive change, response modulation (see the middle layer in  Figure~\ref{fig:ER-5-families}; and the next subsection for details).  
	\item Once a particular family of strategies has been selected, the \textit{implementation} stage involves deciding which specific regulatory tactic (e.g., distraction within attentional deployment) to activate. 
	\item The identification, selection, and implementation are complemented by the \textit{monitoring} system, which decides whether to continue, to stop regulating (as the goal has been achieved), or to switch regulation type following initial implementation (e.g., if the distraction tactic is not working and another strategy needs to be tried out). 
\end{itemize}

For example, difficulties in \textit{identification} would relate to lack of emotional awareness and/or interoceptive awareness of bodily changes related to respective emotion, resulting in the person either over-regulating even subtle signs of emotional states (such as during an anxiety attack), or not being aware of the need to emotion regulate (such as not being aware that one is lashing out at others). 
Issues in \textit{selection} might include limited repertoire of available emotion strategies, whether this is due to habitual selection of maladaptive approaches (such as suppression) or to the lack of knowledge of alternative strategies (e.g., unaware of cognitive reappraisal).  
Difficulties in \textit{implementation} can comprise lack of ability to execute a selected tactic successfully (e.g., attempting cognitive reappraisal might fail if no useful alternative description of the situation is found), or just being overwhelmed by the emotion and aborting the selected technique (e.g., inability to distract oneself from a hurtful memory). 
Finally, failure in the \textit{monitoring} system can involve switching tactics too early (which might have worked if continued), or not stopping maladaptive regulatory tactics early enough (such as engaging in rumination).

\subsubsection{Emotion regulation strategies and tactics}
The 5 \textit{families} of emotion regulatory strategies that could be implemented are situation selection, situation modification, attentional deployment, cognitive change, response modulation -- see the middle row of labels in Figure~\ref{fig:ER-5-families}. Note that these clearly correspond and are differentiated by the stage in the emotion generation process (situation, attention, appraisal, response) that is being targeted by those strategies. Each family then comprises a range of possible  \textit{strategies} that can be utilised as situation-specific tactics to address the situation that a person faces at a particular stage in emotion generation. 
%
Following \cite{Gross2015PsychologicalProspects} and \cite{Sheppes2015}, we now briefly describe each of the families of strategies using the definitions and examples from prior work for clarity, as well as positioning some of the best established emotion regulation strategies within each of the families (cf., \cite{Aldao2010}).
\begin{itemize}
	\item \textit{Situation selection} includes efforts to change the full emotional experience at the earliest stage, such as not choosing to engage situations that might induce unwanted emotions. Examples include avoiding a mean co-worker, or arranging to see an uplifting movie, or choosing to scroll on one's phone rather than engaging with others at a party. One commonly used strategy for situation selection is \textit{avoidance} where the person is choosing a short-term relief through avoiding potentially emotionally risky activities (such as never disagreeing with colleagues due to fear of dealing with conflicts). If used excessively, however, avoidance is seen as an example of a mal-adaptive emotion regulation strategy, with long-term costs outweighing the short-term benefits. 
	
	\item \textit{Situation modification} refers to attempting to change external features of a situation, with the goal of changing its emotional impact. Examples include filing away a rejection letter rather than keeping it fully in view, or shortening the time spent with a colleague who makes you uncomfortable, or blocking a toxic contact on Twitter. A common situation modification strategy often present across a range of interventions is \textit{problem solving}, which is seen as a conscious effort to change a stressful situation or contain its consequences. 
	
	\item \textit{attentional deployment} moves from trying to modify external features of the situation to modifying early information processing that relies on attention. One common strategy is \textit{distraction}, which redirects attention within a given situation (such as focusing on a neutral rather than emotion-eliciting aspect of a movie scene), or shifting internal focus away from the present altogether (e.g., thinking about your afternoon plans rather than engaging in a frustrating meeting). In both such cases, shifting own attention away from emotionally generative aspects of the situation leads to impacts that act very early in the emotion generative process, before the emotional states can gain full strength.	
	
	\item In contrast, \textit{cognitive change} refers to modifying one's own appraisal of the situation in order to alter its emotional impact, i.e., the meaning that we ascribe to the impact of the situation on our goals. Sometimes, cognitive change is applied to an external situation (e.g., "It is okay to fail this interview, I will learn a lot about the industry regardless."). At other times, the cognitive change targets an internal situation (e.g., "My racing heart is not a sign of panic, but rather the excitement before an important presentation.").
	A particularly well researched strategy is \textit{reappraisal}, which targets either the meaning of a potentially emotion-eliciting situation (e.g., "Although my friend is taking a long time to respond to a text message, it must be because she is busy and not because she is ignoring me."), or the self-relevance of a potentially emotion-eliciting situation ("This event doesn't directly involve me or anyone I love."). 
	Another well-known strategy is \textit{non-judgemental acceptance} of emotional states, which is based on a present-centred awareness in which thoughts, feelings, and sensations are accepted as they are. Acceptance is often hypothesised as a core component of mindfulness interventions (see, for example, \cite{farb_mindfulness_2014}).

	\item Finally, \textit{response modulation} refers to directly influencing experiential, behavioural, or physiological components of the emotional response after the emotion is well developed. Examples include using alcohol, cigarettes or drugs to alter emotional state, or relying on deep breathing and exercise to change one's physiological responses. A particularly well-researched strategy in this category is \textit{suppression} which refers to ongoing efforts to inhibit one's emotions and emotion-expressive behaviour and which has been long seen as a mal-adaptive response to stressors (cf., \cite{Sheppes2015, Aldao2010}).
\end{itemize}


There is substantial research on the specific strategies behind each of these categories (cf., \cite{Gross2015PsychologicalProspects}). For example, much is known about the differential usage of strategies in different contexts: reappraisal (as part of cognitive  change category) is highly constructive during low intensity and long-term ER; whereas attentional deployment is often more effective for highly emotionally charged situations \cite{Ford2019a}. Similarly, substantial research has focused on the longer term impact of adaptive and maladaptive strategies (cf., \cite{Aldao2010} for a review), such as the associations between habitual suppression and avoidance use and elevated psychopathology.


\subsection{Articulating the theory framework component -- the `whats'} 

The  psychological model outlined in Section 2 provides an excellent structure for thinking systematically about the intertwined processes of emotion \textit{generation} and emotion \textit{regulation}, the range of cognitive processes involved in emotion regulation attempts, and the specific strategies that could be supported by technology. 
Specifically:

    \begin{itemize}
        \item[-]  \textit{Emotion generation} is a result of: people encountering a \textit{specific situation}, consciously or subconsciously \textit{attending to} it (or not), \textit{appraising} its assumed impact (i.e., evaluating in contrast to own goals), and \textit{responding to} it by potential change in perceived emotion.
The processes of emotion generation are the key substrate on which \textit{emotion regulation processes} operate, by affecting the subsequent emotion generation cycles: situation -- attention -- appraisal -- response.

        \item[-] The process of emotion regulation is seen as proceeding in \textit{4 stages}: \textit{identification} (of need to emotion regulate), \textit{strategy selection} (from the 5 families described above), \textit{implementation} (of a particular instance of a strategy within the current context), and \textit{monitoring}  (whether to continue, stop, or change ER strategy).
				In other words, the distinction between identification, selection, implementation and monitoring stages \textit{describes how the emotion regulation process is started, executed, and eventually ceased}.

				\item[-] Finally, the \textit{five families} of \textit{emotion regulation strategies} (situation selection, situation modification, attentional deployment, cognitive change, response modulation) then synthesise the common approaches that people can implement to up-/down- regulate an emotion at a given moment, each corresponding to an appropriate step in the emotion generation model.

    \end{itemize} 




\section{Strategy components: hows, whens, and wheres of emotion regulation support}

The process model provides a framework to think systematically about the approaches one might take to change a person's emotions and a set of clear intervention targets and strategies. This section builds on this theoretical framework to outline how emotion regulation---as an embodied skill---can be trained. For example, what are the intervention approaches that can help a person alter their predominant strategies (especially if these are already highly habitualised)? How can such process be supported externally, i.e., what are the approaches to help people change their emotion regulation practice and how can these be manualised (developed into replicable training practices) enough to turn into generally applicable interventions? 

In what follows, we combine an overview of learning sciences theory with examples of existing non-technological ER interventions from clinical psychology to articulate the strategy part of the framework: unpacking the approaches to  \textit{`how' emotion regulation can be developed} (i.e., experiential vs didactic); and \textit{when and where learning support can be offered} in emotion regulation interventions (i.e., offline vs on-the-spot). 
The overall aim of this framework component is then to help guide designers in thinking about the use of technology in the context of its role in the \textit{skills development} process: i.e., when, where, and how the design \textit{scaffolds participants' learning}---or application of---emotion regulation skills.

\subsection{Emotion regulation as a `skill development'}
\label{sec:learning_sciences}

It is probably intuitively clear that learning to explicitly regulate emotions requires an understanding of what strategies can be used (`knowledge') as well as how to enact those strategies as tactics in a specific context (`performance').  That is, emotion regulation is not \textit{only} a matter of having the sufficient information: telling someone to `calm down and breathe' during an angry outburst is often no more useful then telling a child to `just keep their balance' whilst sending them down a hill on a first-ever bike ride. In other words, knowing what one should or could do is not enough; one has to be able to execute the selected actions successfully.  
%
%
In the terminology of learning sciences, emotion regulation is a \textit{skill}:  \textit{``an ability that allows a goal to be achieved within some domain with an increasing likelihood as a result of practice''}  \cite{Rosenbaum2001}. 

\paragraph{Two approaches to skills development}\quad While many of the learning processes apply to all kinds of skills, traditional models of learning distinguish between two broad categories:  \textit{intellectual} and \textit{performative} skills. By an `intellectual skill' the literature understands a skill whose goal is predominantly \textit{symbolic} (such as playing chess or solving a mathematics problem). In contrast, a `performative skill' is a skill whose goal is \textit{nonsymbolic} (such as playing a musical instrument, cycling, or playing tennis).
One of the key differences between intellectual and performative skills is in how the necessary knowledge---and thus also the instructions used to teach others---can be encoded in words and well-defined procedures:   
Cognitive \textit{'didactive'} approaches, including explanation and worked examples, are often used for developing symbolic skills (e.g., learning the multiplication procedure in math; cf., \cite{anderson1982a}). In contrast, developing nonsymbolic skills often draws heavily on more  \textit{experiential} approaches (e.g., the repeated attempts at forehand shots in tennis, cf., \cite{Wolpert2011}). 

\paragraph{Where does ER fit?}\quad Positioning \textit{emotion regulation} along the symbolic--nonsymbolic dimension is an interesting problem. On one hand, the goal of emotion regulation is clearly nonsymbolic, insofar as it is aimed at altering a bodily response to a particular person-situation context; it is performative. On the other hand, a substantial part of how such nonsymbolic goal is achieved relies on cognitive processes---including versions of cognitive change (e.g., re-appraisal), explicit attentional deployment, or acceptance---which are often symbolic (using words to re-frame our perception of the situation, for example), and thus it also requires intellectual skills development. 
This then has a substantial impact on the process of skill acquisition, and thus also on how we as designers should think about providing technology-enabled support as part of emotion regulation interventions.

\paragraph{How is ER trained in existing interventions?}\quad In the next section, we showcase two ways of addressing this dual nature of emotion regulation, drawing on established intervention programs in clinical psychology for examples of the learning acquisition approaches. This includes interventions that place a stronger focus on the 
\textit{didactic learning} traditionally applied for symbolic skills (Section~\ref{sec:psych_interventions_experiential}); and those that rely on a more direct \textit{experiential involvement and feedback}, traditionally applied to nonsymbolic skills (Section~\ref{sec:psych_interventions_didactic}). 
We note that an analogous tension between relying on didactic vs experiential approaches is also mirrored in most of the technological systems we will review in Section~\ref{sec:framework}.



\subsection{Exemplar ER interventions -- didactic to experiential}
\label{sec:psych_interventions_didactic}
We selected the two exemplars--- Emotion Regulation Therapy (ERT) \cite{mennin2014emotion} and Unified Protocol for Transdiagnostic Treatment of Emotional Disorders (UP) \cite{Sakiris2019,Ellard2010}---as two cutting-edge, evidence-based interventions that draw on traditional therapy approaches (such as CBT) while positioning emotion regulation techniques at the core of the programme \footnote{We note that emotion regulation approaches are also taught in an analogous way in many other commonly used therapies, including Cognitive Behavioural Therapy (CBT), Dialectical behaviour therapy (DBT), Acceptance and Commitment Therapy (ACT), and have been suggested as one of the few known trans-diagnostic treatments across mental health disorders (see \cite{Sloan2017a,Ehrenreich-May2012,Cludius2020,Sakiris2019} for some of the latest reviews)}.  
%
%

\paragraph{What is taught?}  ERT and UP are multi-week programs, with a wide range of interdependent skills taught.
For example, ERT includes lessons on emotion awareness, self-monitoring and cue awareness, mindful attention, emotional acceptance (reducing rumination/avoidance), cognitive distancing, a range of reappraisal techniques (realistic, positive, compassionate), as well as including exposure therapy where available (i.e., where the fear stimulus is available within therapy sessions and difficult to avoid, such as specific phobias) --- see \cite{mennin2014emotion} for details. 
Each of these techniques has a number of associated exercises, often relying on metaphors and psychoeducation.  
In the context of Gross' model, the techniques taught span all 5 families of emotion regulatory strategies (situation selection, situation modification, attentional deployment, cognitive change, response modulation), as well as all the regulatory stages (identification, selection, implementation, and monitoring).

\paragraph{How are these skills taught?} These interventions predominantly draw on the didactic, symbolic skills development approach, which is then complemented with experiential practice where possible: starting from cognitive psychoeducation---ideally leading to a novel insight for the learner---that is followed up by structured practice within the therapeutic session (e.g., on recalled memories) and then expected to generalise through (often unscaffolded) practice in daily life. 
For example, the general structure of any treatment session within the UP protocol (cf., \cite[p. xxiv]{May2018-book} showcases the common progression and skills development paradigm: The session starts by \textit{reviewing homework assignments} the clients received last time; or doing them in the session by \qqq{generating evocative events and interactions from the week} if the patient has not filled the homework out previously. The therapist then \textit{introduces a new skill} through psychoeducation (describing how the skills work, potentially including carefully crafted metaphors). They then \textit{practice} the new skills in session (on a neutral hypothetical example); continue the practice by applying it to patient's own emotional expression\footnote{The guidebook highlights the importance of emotion generation as part of the session -- "using as many personally relevant examples and activities to reinforce skill knowledge, including the liberal evocation of emotion in session, is highly preferred. Notably there may be some modules where such personalized practice occurs in a separate session from skill introduction."}. Finally, the therapist \textit{assigns home learning assignments}, with the aim to \qqq{allow the clients to apply concepts to real-life experiences outside of the therapy context, which helps generalise skills}. The expectation is that the client will be able to use and apply skills learned in therapy to their life through progressive practice.

\paragraph{Existing design challenges} Enabling the transfer of skill from in-session practice to everyday use is seen as one of the key challenges in existing models, and it is understood that different mechanisms and approaches are necessary to support the learners in these two settings.
For example, ERT explicitly distinguishes between `\textit{offline}' and `\textit{on-the-spot}' versions of taught skills and the associated scaffolding: this corresponds to the notion of developing skills during therapy sessions (`offline'), which relies on psychoeducation and recalling emotional moments from the past with the support from the therapist; and the subsequent attempts to apply these skills in everyday settings where no therapist support is available (`on-the-spot'). 
The distinction between `offline' (training) and `on-the-spot' (in-the-moment) learning approaches---and the associated issue of \textit{transfer} due to lack of mechanisms for on-the-spot support---seems common across many of the existing in-person therapies and programs. 


\subsection{Exemplar ER interventions -- experiential to didactic} 
\label{sec:psych_interventions_experiential}

We selected two examples---Mindfulness-Based Stress Reduction (MBSR) \cite{kabat-zinn_full_2009} and Mindful Awareness in Body-Oriented Therapy (MABT)  \cite{price_interoceptive_2018}---as two body-oriented and evidence based interventions that position emotion regulation techniques, specifically stress reduction, at the core of a therapeutic approach. 
Both styles of intervention target improved ER through modification of perception of physiological signals from the body and related cognitive attributions. 
Although there have been many critics of mindfulness in a western therapeutic context due to the multiplicity of meanings and practices, what is core to these approaches is a focus of attention on body-oriented and felt experience, typically through a focus on the breath, inner body sensations (interoceptive awareness) and/or other bodily processes or sensations, combined with non-judgemental awareness of feelings.   

\paragraph {What is taught?}
In MBSR and MABT interventions, somatically-based and/or meditative style training is used to improve awareness and tolerance of bodily sensations, typically by practicing focused attention on the breath, gentle movement and/or body-scans. The goal is to cultivate attentional skills and improve awareness of the current state of the body and felt experience, and develop a nonreactive awareness of those feelings and sensations as a means to improve ER and well-being \cite{kabat-zinn_full_2009, price_interoceptive_2018}. Both interventions typically proceed through multi-week practice-based sessions focusing on enactment of body-oriented techniques and non-judgemental assessment of those experiences. These approaches primarily support ER skills development around enhanced awareness, attentional deployment, cognitive reappraisal through acceptance and compassion for one’s felt experiences, and may be used in everyday life for response modulation.
In the context of Gross' model, the techniques taught thus predominantly focus on attentional deployment, cognitive change, and response modulation as well as all the regulatory stages (identification, selection, implementation, and monitoring).

\paragraph{How are skills taught?}
These interventions draw heavily on experiential, non-symbolic skills development approaches where the focus is on developing body-based skills through repeated cycles of coached practice. Interventions are structured around staged attentional and interoceptive exercises, in individual or group therapeutic sessions, in order to develop more advanced performance skills over time. In session practice is interspersed with psychoeducational techniques and followed up with additional home-based practice. For example, MBSR focuses on group members being coached in real time on practising a range of experiential techniques including breath-focused meditation, body-scans, mindful eating, walking and yoga/stretching. Layered onto these performative skills development exercises are reminders to cultivate non-judgemental acceptance of whatever physical and emotional sensations arise during practice. A variation called MCBT (Mindful Cognitive Behavioural Therapy), builds on the MBSR multi-week group program with the addition of individual CBT-based sessions, designed to help clients use mindful awareness to reduce negative thought patterns (e.g. associated with depression) \cite{teasdale_prevention_2000}.
Practice sessions are often staged to support the development of skills slowly over time, scaffolded through real-time feedback and/or coaching, psychoeducational discussions, and sharing of experiences either individually with the therapist or in a group therapy setting.


\paragraph{Existing design challenges}
There are two key challenges to body-based emotion regulation models used in ER interventions. One challenge is that experiential or embodied skills development focuses on learning to be aware of, attentive to and able to modify different physiological and neurological processes. Since these  processes are internal body states, they are not easily perceived nor articulated and are not directly observable to an external party (e.g. therapist), making real-time, accurate external feedback and/or self-reflection both difficult and highly subjective; it is often difficult for learners to know when they are `getting it right'. The other challenge is---again---the lack of support to help learners transfer trained skills into everyday life.  Most interventions involve a homework style component in which the client practices what they have learned in sessions, however there is still a gap related to support for learners to transfer embodied skills, such as mindful nonjudgmental attention or interoceptive awareness, into moments of heightened emotional intensity that may occur in everyday life. 
Such approaches thus highlight an analogous difference between \textit{offline} vs \textit{on-the-spot} intervention supports as discussed in the ERT and UP interventions in the previous section. 
\subsection{Articulating the strategy framework component -- hows, whens, wheres} 
\label{sec:conceptualDim_articulation}

\begin{wrapfigure}{r}{0.4\textwidth}
  \begin{center}
    \includegraphics[width=0.38\textwidth]{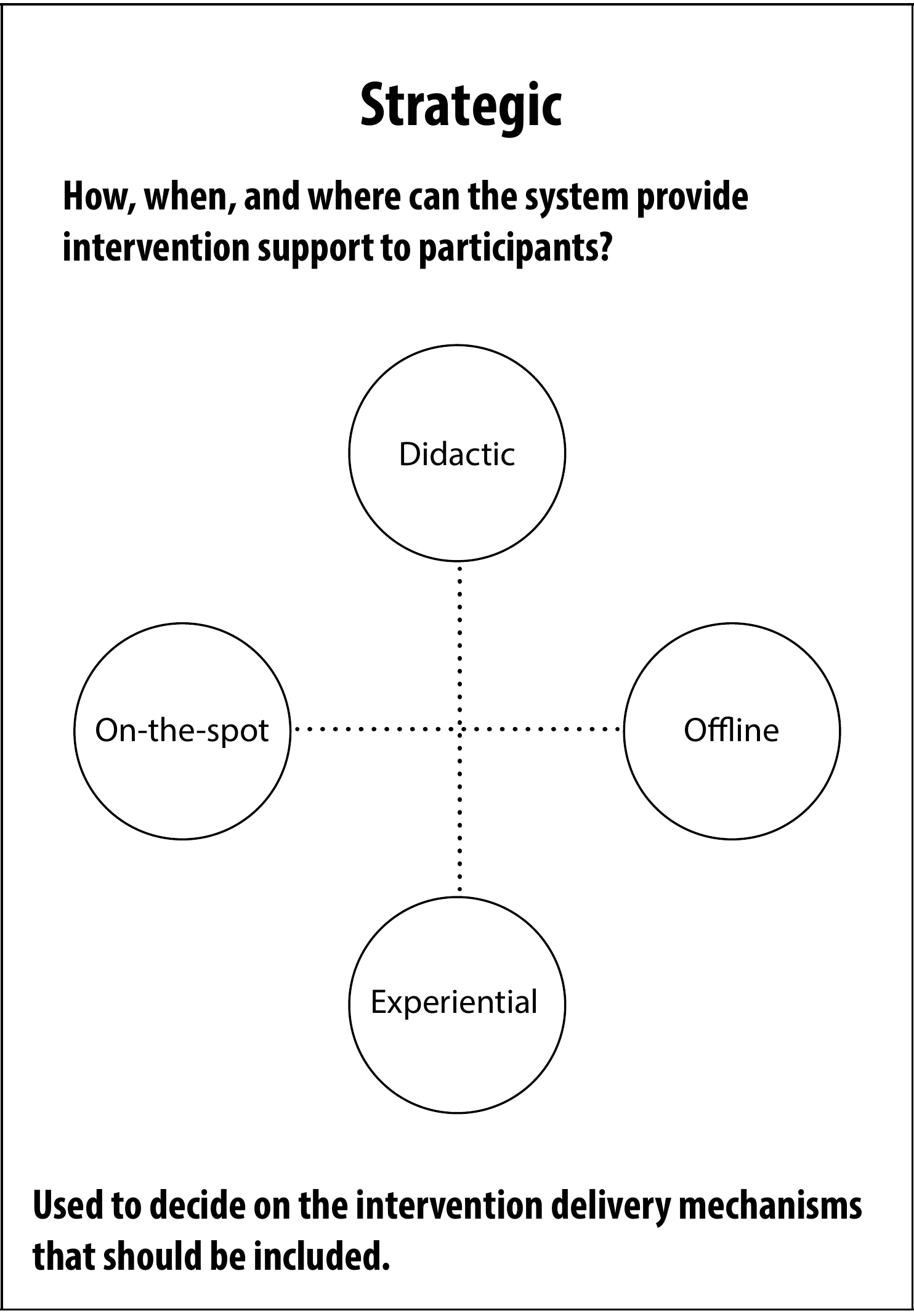}
  \end{center}
  \caption{Strategic framework component}
	\label{fig:strategy-callout}
\end{wrapfigure}

The review above highlights how emotion regulation interventions are best seen as supporting a person in developing an \textit{embodied skill}. 
Drawing on these observations the strategic framework aims to encourage HCI designers to think about the structure of the designed intervention first in the context of the \textit{delivery mechanisms} that digital technology can bring into the intervention process: i.e., \textit{how, when, and where the design is supposed to scaffold participants' learning} -- see Figure~\ref{fig:strategy-callout}. 

\begin{itemize}
    \item[-]  The \textit{hows} part of the framework distinguishes \textit{the type of skills acquisition approach} an intervention component supports: distinguishing between \textit{didactic learning} (such as the psychoeducation approaches used in UP/ERT and delivered e.g., by the therapist or through reminders/homesheets)  and  \textit{experiential learning} (such as the trial\&feedback approach commonly used in mindfulness and delivered e.g., through biofeedback). 

    \item[-]  The \textit{whens/wheres} part of the framework then helps the designers think about \textit{timing and contexts in which intervention support is provided} through technology: distinguishing between  \textit{`offline' training} (such as intervention support in a therapist office or an online module); and \textit{`on-the-spot' learning} (such as situated support during a stressful real-world situation).
\end{itemize}

\begin{figure}[t]
    \centering
    \includegraphics[width=0.98\textwidth]{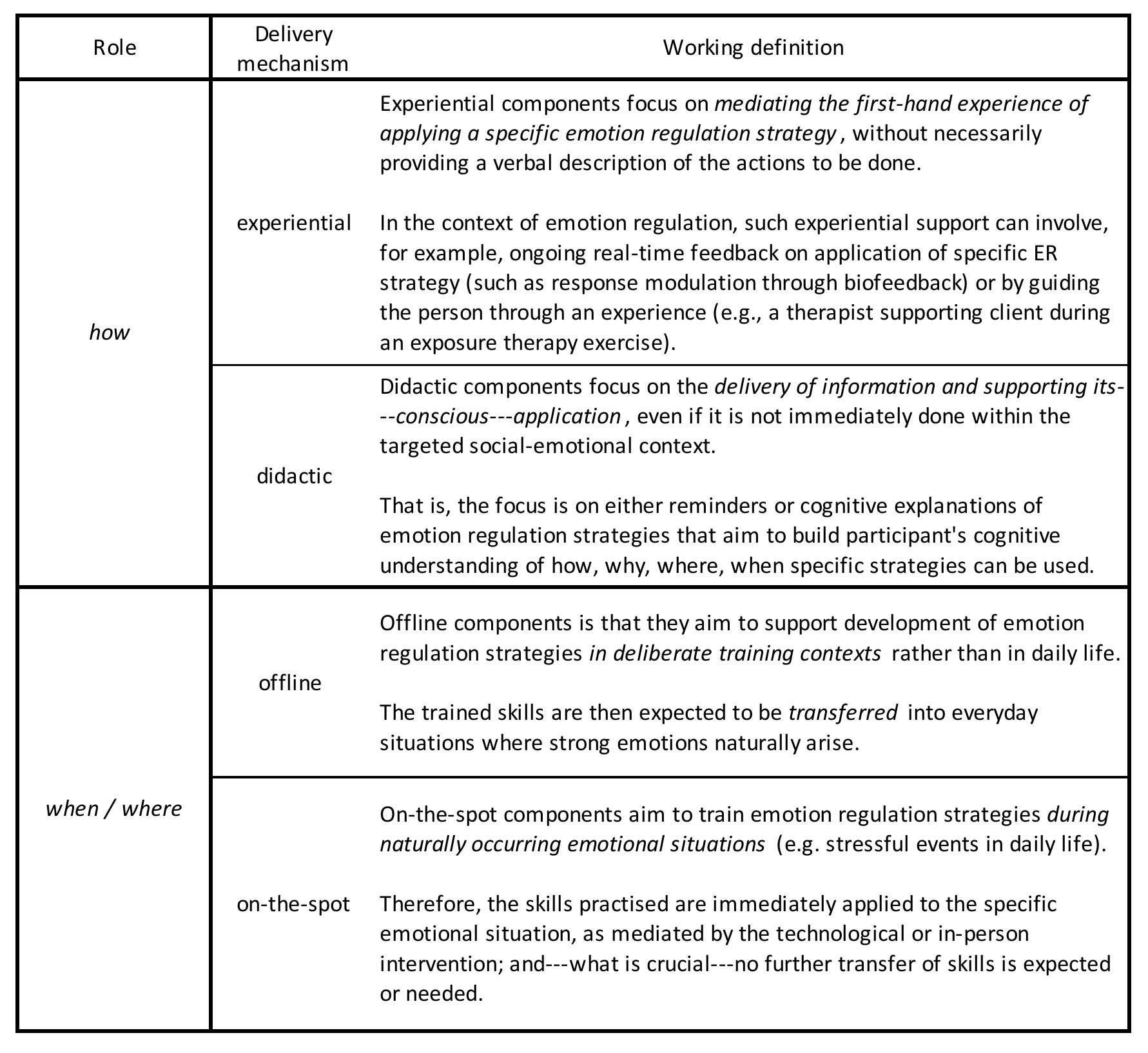}
    \caption{ Definitions table for each of the delivery mechanism components: experiential vs didactic; offline vs on-the-spot.}
    \label{fig:mechanisms-table}
\end{figure}

Figure~\ref{fig:mechanisms-table} outlines each of the delivery mechanisms in more detail, including the definitions used to categorise existing HCI work in the scoping review presented in the next section. As argued in more depth in Section 5, we propose that this conceptual structure can enable modular thinking about intervention design, provides a framework for the community to develop an empirically grounded understanding of `what works' for each type of delivery mechanism (possibly across a range of technologies), as well as promoting cross-project inspiration across HCI and psychology research teams. 

We note that such modularity is already visible in all of the---non-technological---intervention programmes described in this section. These combine a range of complementary delivery mechanism components: for example, including delivery of \textit{didactic} (psychoeducation \& scaffolded application in session) and \textit{experiential} learning opportunities (repeated practice and in-situ skills generalisation); with a range of support available to the learners as part of \textit{`offline'} training (guided meditation) and \textit{`on-the-spot'} application contexts (e.g., home-learning assignments).
We expect an analogous situation also for any technology-enabled interventions:  for example, a game-based bio-feedback intervention (such as \cite{Knox2011} reviewed in the next section) can be seen as utilising `offline' delivery mechanism through the game (which is used to generate emotions outside of everyday situations), `experiential' mechanism (by involving the biofeedback loop), as well as perhaps a `didactic' component (if, for example, the participants first engage in psychoeducation before the game starts).

\vfill . \pagebreak

\section{Practical component: scoping review of HCI work}
\label{sec:framework}



The role of the practical part of the framework is to support  designers in answering questions such as \textit{How can we provide an experiential/didactic/offline/on-the-spot support to those using my interventions?} \textit{What are the established technological approaches to do so, and where is there still an opportunity for innovation?}  
In other words, our aim is to provide designers with a synthesis of the existing work on emotion regulation technologies in HCI to date into commonly used \textit{intervention design components} across the selected delivery mechanisms, help guide practical decisions on possible implementations of these, as well as help identify areas where current solutions need improvement or are missing.

The descriptions that follow are based on a scoping review (cf., \cite{Munn2018_ScopingReview}) including papers published in HCI between January 1 2009 and December 8 2021, as primarily indexed across ACM digital library and IEEE; and complemented with an additional Google Scholar search.\footnote{
The authors (i) will update the review for camera ready version; as well as  (ii) maintain an up-to-date database of papers for at least 5 years after the publication on OSF (as part of Slovak’s FLF fellowship).
}
We use this scoping review of recent HCI literature in two ways: 
\begin{enumerate} 
	\item To develop a `dictionary' of commonly used \textit{intervention design components}, as identified in existing work for each of the delivery mechanisms  (i.e., what are the design choices researchers have tended to use so far to provide experiential, didactic, offline, or on-the-spot support?); as well as to provide an overview of \textit{the range of the ER targets} within the current work (i.e., which of the 4 ER stages/5 families of ER strategies have been supported so far, and which remain under-studied?).
	\item To outline how these individual design components have been \textit{combined into interventions} (i.e., which combinations of the intervention design components have been explored already in the literature? What are the remaining gaps?). 
\end{enumerate}

\begin{figure}
  \centering
	\includegraphics[angle=0, width=0.9\textwidth]{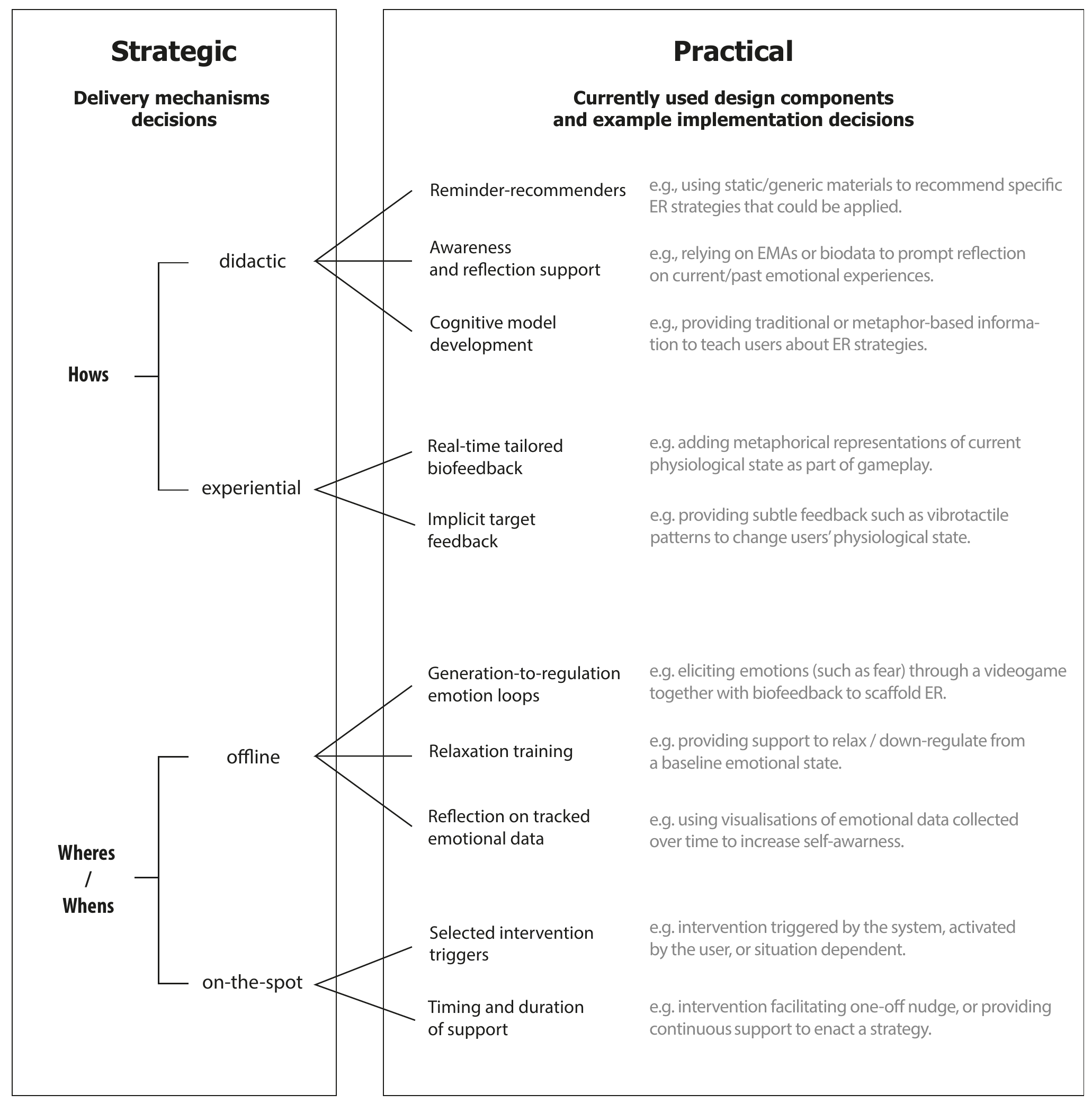}
	\caption{Overview of the connections between strategic and practical parts of the framework.}
	\label{fig:sec3-4-overview}
\end{figure}

Figure~\ref{fig:sec3-4-overview} visualises how the strategic and practical parts of the framework connect and provides an overview of the main \textit{intervention design components} identified for each of the four delivery mechanisms.  

%

%


\subsection{Scoping review methodology}
We conducted a literature review that targeted full papers and notes as well as high quality works-in-progress that were published in top venues in HCI. We targeted the ACM digital library, IEEE databases and used Google Scholar to search for papers published between January 1 2009 and December 8 2021. Our keyword search criteria were “interactive technology” and one or more of “emotion regulation” or “self-regulation” or “stress” or “stress-regulation”. The initial search resulted in 5,574 papers, out of which 36 papers fit all our criteria: peer reviewed papers addressing human emotion-regulation, involving interactive technology, including user study, identified one or more intervention mechanisms, and where the end-goal of the research was an emotion regulation intervention (as opposed to say, e.g., an emotion detection algorithm). 
Appendix~1 provides the detailed information about the selection and filtering process. 

Following from the theoretical review in Sections 2 and 3, we have used an inductive and iterative approach to thematic analysis to create a coding scheme and identify intervention components within each system that supported offline vs on-the-spot training, and didactic vs experiential learning. The analysis phase consisted of two of the researchers working inductively to iteratively develop initial categories and coding rules for each of the four delivery mechanisms. This was done during individual and shared analysis sessions over a small subset of papers. Once categories were defined, the researchers then individually reviewed and coded about 25\% of the papers across the two dimensions, together reconciling, revising and finalizing descriptions of categories and coding rules for each of the categories. This codebook was finalised after the researchers had read and analyzed about half the papers, at which point they had well-defined dimension categories and stable coding schemes for each dimension. 
 These observations form the core contribution of this section and are outlined below. 

As part of the initial approach, the researchers also coded the systems along a more traditional six dimensions, selected a priori:  (1) type of technology; (2) stage of research to, e.g., distinguish early HCI explorations of materials from more developed systems; (3) topical focus of HCI research; (4) intervention mechanism underpinning the designed system; (5) level of intervention mechanism specificity; and (6) intended use-case (skills development vs continuous support). These dimensions have been derived from prior research and similar HCI review work, as well as the theoretical grounding from psychology (i.e., focus on theory of change, and learning vs on-going support). These are very briefly summarised in the next subsection and reported in full in Appendix 1. 

\subsection{Overview of HCI work}
\label{sec:HCI_overview}

The papers in our sample showcase a wide variety of HCI research in the design space of creating and evaluating emotion regulation interventions, with the main computing platforms being mobile devices (16 papers), PCs/laptops (11 papers), followed by a range of other technologies (e.g., smart watches (4), bespoke embedded computing prototypes (10)); many supplemented with biosensors and non-traditional output devices (e.g. haptics, ambient lighting). 
 
As is traditional in HCI research (cf., e.g,. \cite{Sanches2019}), the contribution of many of the papers focused on early stage explorations of innovative technologies in design space, presenting user studies on proof-of-concept devices to examine whether a particular approach/class of technologies was viable and warranted future research. Specifically, 13 papers were coded as \textit{early stage research}, focusing on the design of prototypes and formative evaluations; 14 papers were coded as \textit{mid-stage research}, focusing on iterative design and evaluation of more complex and/or robust prototypes with formative or summative evaluations; finally, only 9 papers were coded as \textit{late stage research} involving hi-fidelity, robust research systems that were ready for widespread deployment and summative evaluations in lab or field studies.
Many of the more exploratory systems did not fully specify the assumed theory-of-change (cf., \cite{DeSilva2014}), i.e., outlining how the proposed design decisions are likely to lead to users' improved emotion regulation, with about half of the systems being coded as including \textit{low} (15 papers) or \textit{medium} (12 papers) specificity of the theory of change.   

In terms of interaction design, most research papers focused on utilising biofeedback components (14 papers), physiological synchronisation, i.e., providing users with external input that mimics their own biosignals such as `false heart-rate feedback' (7 papers); reminders and recommendations of specific ER strategies (6 papers); awareness systems that help monitor and/or track individual emotions (5 papers);  and a smattering of other bespoke systems (4 papers). 
Interestingly, a unifying characteristic across all these different approaches was the lack of focus on intervention transfer beyond the effects during technology use: the strong majority of the papers in the current data set (n = 31) assume that the developed systems have to provide an on-going support for the intervention to be effective; and thus the effects would disappear if the system (such as wearable feedback) was removed. In contrast, only a handful (n = 5) of interventions were specifically designed to reduce end-user dependence on the intervention to enable the end-user to apply emotion regulation strategies on their own (cf., \cite{Antle2019,Antle2018,Scholten2016,Zafar2017,Slovak2016}). 

\subsection{Which intervention design components are used in each of the delivery mechanisms?} 
\label{sec:components}

\begin{figure}[t]
    \centering
    \includegraphics[width=.9\textwidth]{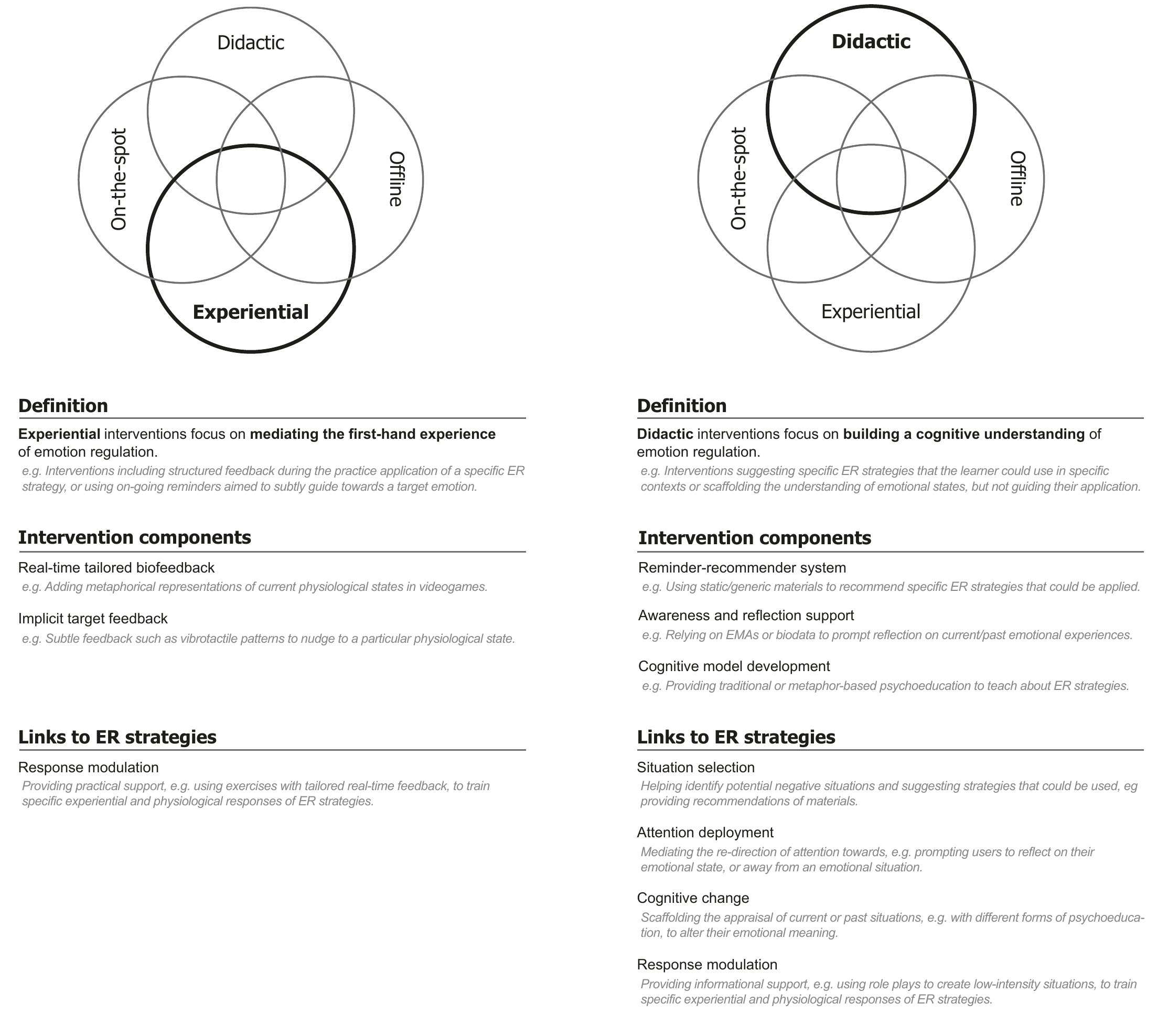}
    \caption{Overview of the experiential and didactic intervention components}
    \label{fig:BU-vs-TD}
\end{figure}
\begin{figure}[t]
    \centering
    \includegraphics[width=.9\textwidth]{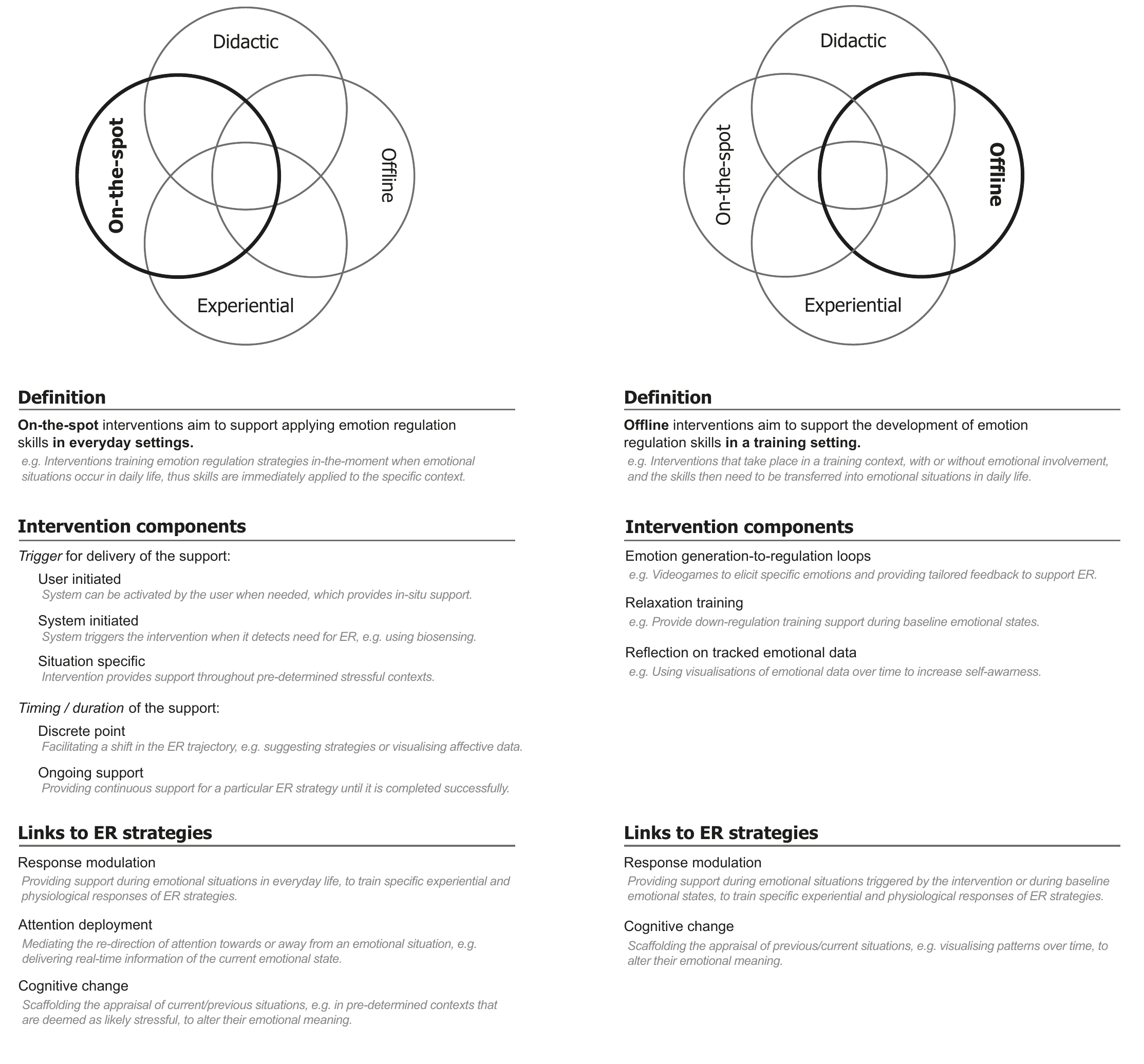}
    \caption{Overview of offline and on-the-spot intervention components}
    \label{fig:ITM-vs-OOC}
\end{figure}

Our next step was to identify and categorise the \textit{design components} that have been used in existing work so far. The aim was to articulate the established---or at least attempted---approaches to supporting each of the four possible delivery mechanisms (e.g., what are all the components that have been used to support `experiential' learning in prior HCI work?). 
%
%
In particular, each delivery mechanism---didactic, experiential, offline, on-the-spot---is briefly characterised in terms of the ER targets, narrative overview of how we saw the respective components used in the current dataset, and the categories of components identified. Figures \ref{fig:BU-vs-TD} and \ref{fig:ITM-vs-OOC} provide the overview of the contrasts between experiential vs didactic; and offline vs on-the-spot components (on pages \pageref{fig:BU-vs-TD} and \pageref{fig:ITM-vs-OOC}, respectively). 
 

\subsubsection{Experiential intervention components}
\label{sec:component-BU}
\paragraph{ER targets} The main characteristic of intervention components coded as experiential is their \textbf{focus on mediating the first-hand experience of applying a specific emotion regulation strategy}, without necessarily providing a verbal description of the actions to be done -- cf., the section on performative skills development. In the context of emotion regulation, such experiential support can involve, for example, ongoing real-time feedback on application of specific ER strategy (such as response modulation through biofeedback) or by guiding the person through an experience (e.g., a therapist supporting a client during an exposure therapy exercise).

\paragraph{How have these been used in HCI work?} Predominantly, experiential components attempt to facilitate the learner's on-going emotion regulation process, such as when the learner would otherwise struggle to apply the emotion regulation strategy successfully. Although more than half of the reviewed papers contained some experiential components (23 papers in total), the range of mechanisms used was limited to the two main categories outlined below. The first provided the user with a learning environment that included structured feedback during a practice application of a given ER strategy, such as embedding biofeedback into video games (e.g., \cite{Wang2018, Mandryk2013}). This approach was a goal-driven or purposeful, conscious approach and scaffolded experience(s) that involved explicit control of emotion regulation. The second approach used on-going non-lingual reminders that aimed to subtly guide the participant toward a target emotional state, such as through vibrotactile patterns (e.g., \cite{Miri2020, Costa2019, Paredes2018, Slovak2018}) or audiovisual cues (e.g., \cite{Ghandeharioun2017}). In this approach the goal was not communicated directly to the participant and involved scaffolding implicit emotion regulation experiences through unconscious or automatic processes.

\paragraph{Interaction design patterns} The two main interaction designs used across the 23 papers---\textit{real-time tailored biofeedback} and \textit{implicit target feedback}--- differed substantially in how, when, and through which psychological mechanisms users' emotion regulation was to be supported. 
%
The \textit{real-time tailored biofeedback} components provided the user with information about their current physiological state(s) in a closed (bio-)feedback loop, which then helps to learn and/or practice ER through operant conditioning \cite{Antle2018,Antle2019,Knox2011,Liang2018,Lobel2016,Mandryk2013,Parnandi2017,Parnandi2018,Scholten2016,Wang2018,Yu2018,Zafar2017,Jiang2021}. This was predominantly implemented through video games that map one's physiological response to the system in real time using metaphorical representations, either subtly changing the game mechanics \cite{Antle2018, Antle2019,Knox2011,Lobel2016,Scholten2016,Parnandi2017, Parnandi2018,Wang2018, Zafar2017} or using graphic overlays that make the gameplay more difficult if the player is not calm enough \cite{Mandryk2013,Lobel2016,Wang2018}. For example, Mandryk et. al \cite{Mandryk2013} inserted additional graphics overlay (e.g., mud in a racing game) to gameplay as the player's stress increased.
In contrast, \textit{implicit target feedback} components aimed to subconsciously or consciously nudge the users' physiological states towards a particular target (often relaxation or calmed down state) \cite{Balters2020,Carlier2019,Costa2019,Ghandeharioun2017,Miri2020,Morajevi2011,Paredes2018,Slovak2018,Wells2012}. Predominantly, such components utilise vibrotactile interaction, either as a guide for the users' controllable physiological processes, such as their breathing rate \cite{Ghandeharioun2017,Morajevi2011,Paredes2018,Miri2020,Balters2020,Wells2012}, by drawing on the false-heart rate feedback literature (e.g., \cite{Costa2019}), or by relying on attentional deployment strategies elicited by the haptic interaction \cite{Slovak2018}. For example, Miri et. al's \cite{Balters2020} system provides haptic feedback to facilitate calm breathing patterns of drivers during a car commute.

\paragraph{Links to the ER process model}
\noindent The range of ER techniques supported by experiential components was very limited: nearly all of the implicit target and biofeedback components were supporting ER strategies belonging to the \textit{response modulation} family, often through affecting the breathing rate. The exceptions to this pattern were bio-feedback systems embedded in traditional therapeutic techniques (mostly CBT -- see \cite{Knox2011,Lobel2016,Antle2019}), which would have introduced other families of ER strategies through different components. Even there, however, the core bio-feedback loop relied mostly on breathing-based exercises. Additionally, many of the systems---especially in the implicit target feedback category---did not explicitly aim to promote skills development; the system would need to be present at all times when the user needed to down-regulate. 


\subsubsection{Didactic intervention components}
\label{sec:component-TD}
\paragraph{ER targets} The main characteristic of intervention components coded as didactic is their focus on the \textbf{delivery of information and supporting its---conscious---application}, even if it is not immediately done within the targeted social-emotional context. That is, the focus is on either reminders or cognitive explanations of emotion regulation strategies that aim to build participants' cognitive understanding of how, why, where, and when specific strategies can be used.  

\paragraph{How have these been used in HCI work?} Didactic intervention components rely on active cognitive processing to help develop new (or plan to apply already known) emotion regulation strategies. Many existing examples rely on technology to provide information suggesting specific ER strategies that can be applied to a particular emotion regulation need (e.g. \cite{Morajevi2012, Paredes2014, Diaz2018, Fage2019}; or scaffold understanding of own emotional states to enhance self-awareness and monitoring over time, for instance, via self-reports (e.g. \cite{Huang2015}) or sensing technology (e.g. \cite{Morajevi2012}). What is important to note is that these intervention components do not scaffold the experiential aspects of emotion regulation (e.g., application of the suggested interventions), although they might be explicitly describing the assumed cognitive steps that the learner \textit{should take} once the strategy is to be applied. Finally, some interventions rely on traditional psychoeducation approaches to explain, suggest, and support development of new emotion regulatory approaches (e.g., by providing verbal explanations of how to execute emotion regulation strategies) as well as including role plays to provide low-intensity situations to practice on (e.g., \cite{Knox2011,Pina2014,Scholten2016}).

\paragraph{Interaction design patterns} We saw three main interaction design approaches---\textit{reminder-recommender systems}, \textit{awareness and reflection support}, and \textit{cognitive models development}---which differed in when, how, and what kind of information was offered to the users. 
	First, \textit{reminder-recommender components} provide the users with in-the-moment suggestions around specific ER strategies that could be used \cite{Diaz2018,Fage2019,Huang2015,Paredes2014,Pina2014,Smyth2016}. Often, the implementation included providing step-by-step instructions on how to enact a particular ER strategy in the form of static materials (i.e., general description of how to `do' cognitive reappraisal or deep breathing, not tailored to any current experience), or by providing links to external materials. For example, Smyth et. al. \cite{Smyth2016} delivered three reminder prompts daily (e.g., “Recall that even  a  few  diaphragmatic  breaths  can  reduce  stress;  if  your current  context  allows  for  it,  take  5  mindful  breaths.”) at a semi-random time, through a mobile phone notification interface.   
	Second, the \textit{awareness and reflection} interaction mechanisms do not suggest specific activities but rather rely on prompting the users to reflect on their current emotional state and/or expect them to analyse a (retrospective) data log of one's emotional experiences \cite{Bakker2018,Huang2015,Moraveji2012,Kocielnik2013,Smyth2016,Wang2019}. The reflection support is often reliant on users' self-report through ecological momentary assessment (EMA) \cite{Bakker2018,Huang2015,Smyth2016}, or automated system tracking relying on biosensing data \cite{Kocielnik2013, Wang2019}. For example, Bakker et. al. \cite{Bakker2018} used a bespoke app to prompt users to collect their emotional states  (12 questions once per day), and then enable them to reflect on any patterns visible in the resulting 'mood diary'. 
	Finally, \textit{cognitive model development} components do not scaffold any ongoing engagements, but rather  provide users with psychoeducation (and potentially role-play) support to develop an understanding of new emotion regulation strategies \cite{Antle2018,Antle2019,Knox2011,Pina2014,Scholten2016,Slovak2016}. This can take place in traditional clinical intervention context with a therapist \cite{Pina2014,Knox2011}, as part of embedding the psychoeducation into the interactive experience \cite{Scholten2016, Slovak2016}, or as well-selected metaphors incorporated into an interactive game \cite{Antle2019, Antle2018}. Specifically, these components provide in-depth explanations of how the mind transforms information to arrive at behaviour (e.g. how, why, where, when it is important to use ER skills).


\paragraph{Links to the ER process model} 
\noindent  Didactic components have supported emotion regulation strategies from a range of families, although these were mostly determined by the particular interaction design mechanism. For example, the reminder-recommender system components mostly supported some form of \textit{situation selection} although there were some implementations that aimed at a broader range (e.g., \cite{Pina2014,Paredes2014}, which referred to \textit{cognitive change} and \textit{response modulation} strategies as well). Cognitive models' development components were predominantly relying on traditional CBT psychoeducation (e.g., \cite{Knox2011, Scholten2016}) and thus supported a broad range of strategies similar in scope to the clinical interventions outlined in Section 2, including \textit{attentional deployment, cognitive change, response modulation}; others have centred more on \textit{acceptance} approaches (\cite{Antle2018,Antle2019}. Finally, the Awareness and Reflection systems predominantly aimed at supporting situation selection (by helping learners identify situations that lead to negative outcomes), and potentially also the self-awareness meta-layer. 



\subsubsection{Offline intervention components}
\label{sec:component-offline}
\paragraph{ER targets} 
The main characteristic of offline intervention components is that they aim to support development of emotion regulation strategies \textbf{in deliberate training contexts} rather than in daily life. The trained skills are then expected to be \textit{transferred} into everyday situations where strong emotions naturally arise.

\paragraph{How have these been used in HCI work?} \quad  Offline intervention components can take a number of forms. For instance, systems including offline components can aim to elicit strong emotions using technology to create an artificial context in which to practice ER strategies \cite{Lobel2016, Mandryk2013, Parnandi2017, Parnandi2018, Wang2019, Zafar2017, Scholten2016}. This is predominantly achieved through training contexts involving video games designed to either trigger strong emotions (such as fear in \cite{Lobel2016} or stress in \cite{Zafar2017}) or to illustrate metaphorical representations that scaffold the successful execution of the ER strategy being trained (e.g. \cite{Antle2018, Scholten2016}). In many instances the existing components predominantly rely on forms of biofeedback,  i.e. \textit{experiential components}, to `close' the training loop (cf., Section~\ref{sec:component-BU}).  Alternatively, several interventions use ongoing monitoring of tracked emotions across situations to create an overview that is used in a training context to enable reflection on previously unnoticed patterns, resulting in insights and an eventual increase in emotional awareness over time (\cite{Bakker2018, Kocielnik2013, Huang2015, Wang2019}). Multiple papers that investigate mechanisms that enable offline training also include the use of \textit{didactic} psycho-education (cf., cognitive models development in Section~\ref{sec:component-TD})  to teach participants about when and how emotion regulation strategies are useful and when to use them in real life \cite{Wells2012, Scholten2016, Knox2011, Smyth2016, Pina2014}. The psycho-active information is delivered to the users predominantly through non-technologically mediated ways such as in-person therapy sessions or training \cite{Wells2012, Knox2011, Pina2014, Smyth2016}, with \cite{Scholten2016} being the only exception where the psychoeducation has been directly incorporated into the technologically mediated intervention (computer game). Overall, the psychoeducation approaches are designed to supplement the main technology-enabled component (e.g. to support the teachings from biofeedback interventions \cite{Knox2011, Scholten2016, Antle2019}). 

\paragraph{Interaction design patterns}\quad  The existing approaches to offline design components differ primarily in \textit{if/how the emotion generation process is supported through technological mediation}. 
%
The first type of components aims to \textit{both facilitate the emotion generation and the associated emotion regulation}. These are mostly intervention components utilising the ability of video games to elicit specific emotions (e.g. anxiety, stress) but then also facilitate feedback on the emotion regulation of the player (e.g. using biofeedback about the emotional state of the player). This interaction mechanism enables `in-vivo' ER practice as part of the game experience, with the mostly negative emotion either purposefully elicited through game design (e.g. \cite{Lobel2016, Scholten2016}) or generated through gameplay of commercially available games which are inherently stressful (e.g.\cite{ Mandryk2013, Wang2018}). 
In the second type of components, the technological systems do not attempt to elicit heightened emotion but instead \textit{provide relaxation training support} during baseline emotional states \cite{Knox2011, Antle2019, Antle2018, Carlier2019}. The focus here is often on utilising technology-enabled feedback mechanisms, again typically biofeedback to ease and/or deepen the relaxation competence of the player (e.g., \cite{Antle2018, Liang2018}). 
Finally, there are offline components that neither generate emotion nor scaffold the resulting emotion regulation. Instead, the focus is on the users' reflection on emotion data by \textit{visualising patterns over time}. In these instances the system draws on a history of tracked emotional data (and relevant context) with the expectation that visualising this data to the user will lead to insight and increased self-awareness \cite{Bakker2018, Kocielnik2013, Huang2015, Wang2019}. 
Such systems may employ mechanisms that automatically collected physiological data (e.g. \cite{Wang2019}), user-entered data, such as a short mood questionnaire \cite{Bakker2018};  and/or a location-based emotion assessment \cite{Huang2018} alongside various representational forms for display. For example, Bakker et. al. allow the users to inspect a 'mood diary' to extract patterns, increase self-awareness of their emotions, and potentially change their future behaviour \cite{Bakker2018}. 

\smallskip

\paragraph{Links to the process model of ER}
The majority of game-based interventions are predominantly focused on training \textit{response modulation} techniques (e.g. control over aspects of breathing). Conversely, emotion awareness interventions are not grounded in any specific theory of change, instead relying on the notion that improved awareness of emotions will result in better ER. Interventions include a secondary psychoeducation element to focus on \textit{cognitive reappraisal} or simply include explanations and reminders to use the techniques being taught. 


\subsubsection{On-the-spot intervention components}
\label{sec:component-OTS}
\paragraph{ER targets}\quad The main characteristic of intervention components coded as on-the-spot training is that they aim to train emotion regulation strategies \textbf{during naturally occurring emotional situations} (e.g. stressful events in daily life). Therefore, the skills practised are immediately applied to the specific emotional situation, as mediated by the technological or in-person intervention; and---what is crucial---no further transfer of skills is expected or needed. 

\paragraph{How have these been used in HCI work?}\quad Some of the systems including on-the-spot components target moments when a strong emotion arises, such as: during a stressful work situation (e.g. \cite{Miri2020,Yu2018, Morajevi2011}) or whilst driving in a busy city (e.g. \cite{Paredes2018}, \cite{Balters2020}), or everyday feelings of stress or anxiety (\cite{Slovak2018,Fage2019}). Alternatively, some interventions deliver real-time information of the current emotional state (e.g. \cite{Morajevi2012}, \cite{Wang2018}) and some suggest potential ER strategies to be used when a specific emotion is experienced or sensed through the technology (e.g. \cite{Smyth2016}, \cite{Huang2015}).  Many of the papers highlight the focus on investigating mechanisms that enable such on-the-spot delivery of intervention \textit{alongside} other ongoing activities, mostly in combination with \textit{experiential} components, cf., \cite{Moraveji2011,Costa2019,Ghandeharioun2017,Paredes2018,Yu2018,Miri2020,Balters2020,Moraveji2012}; see Section~\ref{sec:component-BU}). Alternatively, other papers use more \textit{didactic} approaches, predominantly in the form of reminders or EMAs (\cite{Bakker2018,Diaz2018,Fage2019,Huang2015,Smyth2016,Moraveji2012,Paredes2014,Pina2014,Wang2019}; see Section~\ref{sec:component-TD}).

\paragraph{Interaction design patterns}
The existing approaches differ in two key ways that shape the interaction designs enabled by the components: these are \textit{what is used as the trigger} for delivery of intervention supports (user initiated, system initiated, situation specific); and \textit{the timing/duration} over which the support is delivered (discrete point or ongoing support). 
We outline each of these below, referencing the systems that utilised such interaction design mechanism(s). 
\smallskip

The differences in approaches to \textit{intervention triggers are in who/what initiated the on-the-spot intervention delivery}: 
For the \textit{user-initiated} components \cite{Fage2019, Huang2015, Liang2018, Slovak2018, Wang2019,Moraveji2012}, the users were expected to notice the need for emotion regulation support and activate the system, with the on-the-spot intervention component facilitating their efforts afterwards. For example, BioFeedback \cite{Liang2018} requires the user to reach out for the interactive fidget spinner, which then provides in-the-moment feedback. 
For the \textit{system-initiated} components \cite{Pina2014,Costa2019,Ghandeharioun2017,Smyth2016,Paredes2014}, it is the system that determines the need for emotion regulation at a particular time---often through bio-sensing---and triggers intervention delivery: \cite{Pina2014} (EDA), \cite{Costa2019} (HR), \cite{Ghandeharioun2017} (breathing rate), \cite{Smyth2016,Paredes2014} (algorithmically based on user's EMA input). For example, \cite{Smyth2016} relies on detecting the user's high stress or negative affect (as part of a prompted EMA), to then provide tailored reminders to use stress management strategies via a mobile app.
Finally, the \textit{situation specific} components provide ongoing support throughout predetermined contexts that are deemed as likely stressful (e.g., driving (\cite{Paredes2018, Balters2020}, or information work (\cite{Moraveji2011,Yu2018}). For example, CalmCommute \cite{Balters2020} provides continuous haptic-guided slow breathing exercises through a modified car-seat cover, while the user is driving. 
\smallskip

The differences in approaches to \textit{timing and duration of the intervention refer to the period over which the support is offered on-the-spot}. For the \textit{discrete point} components, the system provides the user with information that is expected to facilitate a shift in their emotion regulatory trajectory. In present dataset such components relied on either reminding the users of strategies taught in prior psychoeducation modules \cite{Pina2014, Smyth2016}, facilitating awareness of their stress levels \cite{Yu2018, Wang2019}, or providing users with suggested activities to improve emotional state \cite{Huang2015, Paredes2014}. For example, \cite{Pina2014} used visual representation of in-the-moment strategies on glance-able displays positioned in the home to help parents remind themselves of strategies they were taught in prior therapeutic sessions, with the expectations that these strategies could then be put to use (without further scaffolding). 
In contrast, with the \textit{ongoing support} components the users are supported through a particular emotion regulation strategy until the emotion regulation is complete. In existing work, the components predominantly support explicit ER by guiding the users toward slower breathing patterns \cite{Miri2020, Moraveji2012, Paredes2018, Balters2020, Liang2018}, but also by providing cognitive scaffolding \cite{Fage2019} and ongoing attentional deployment intervention \cite{Slovak2018}. Other work relies on implicit ER, where users are expected to subconsciously alter their ongoing behaviour and/or emotions in response to a `target feedback' delivered as haptic interaction (\cite{Costa2019, Choi2020, Ghandeharioun2017}). For example, BoostMeUp \cite{Costa2019} used Apple Watch to give users subtly perceptible haptic `taps' on the wrist which were either 30\% slower or 30\% faster than the participants' baseline heart rate. 

\smallskip

\paragraph{Links to the process model of ER}
As these components are related to the point and type of support offered, rather than the specific content, there are no in-principle connections between the specific interaction design mechanism components (e.g., user-initiated vs system-initiated), and specific emotion regulation families from the Process model, or the overarching stages (identification, selection, implementation, monitoring). For example, while ongoing support components are now predominantly used for \textit{response modulation} (e.g., through deep breathing or false heart-rate feedback), there are no principled reasons why a similar approach could not be used to support other families (e.g., attentional deployment, or  cognitive reappraisal); or across the full gamut of the overarching stages. 
\medskip


\subsection{How are the identified design components combined in existing interventions?} 
\label{sec:clusters}


\begin{figure}[t]
    \centering
    \includegraphics[width=0.9\textwidth]{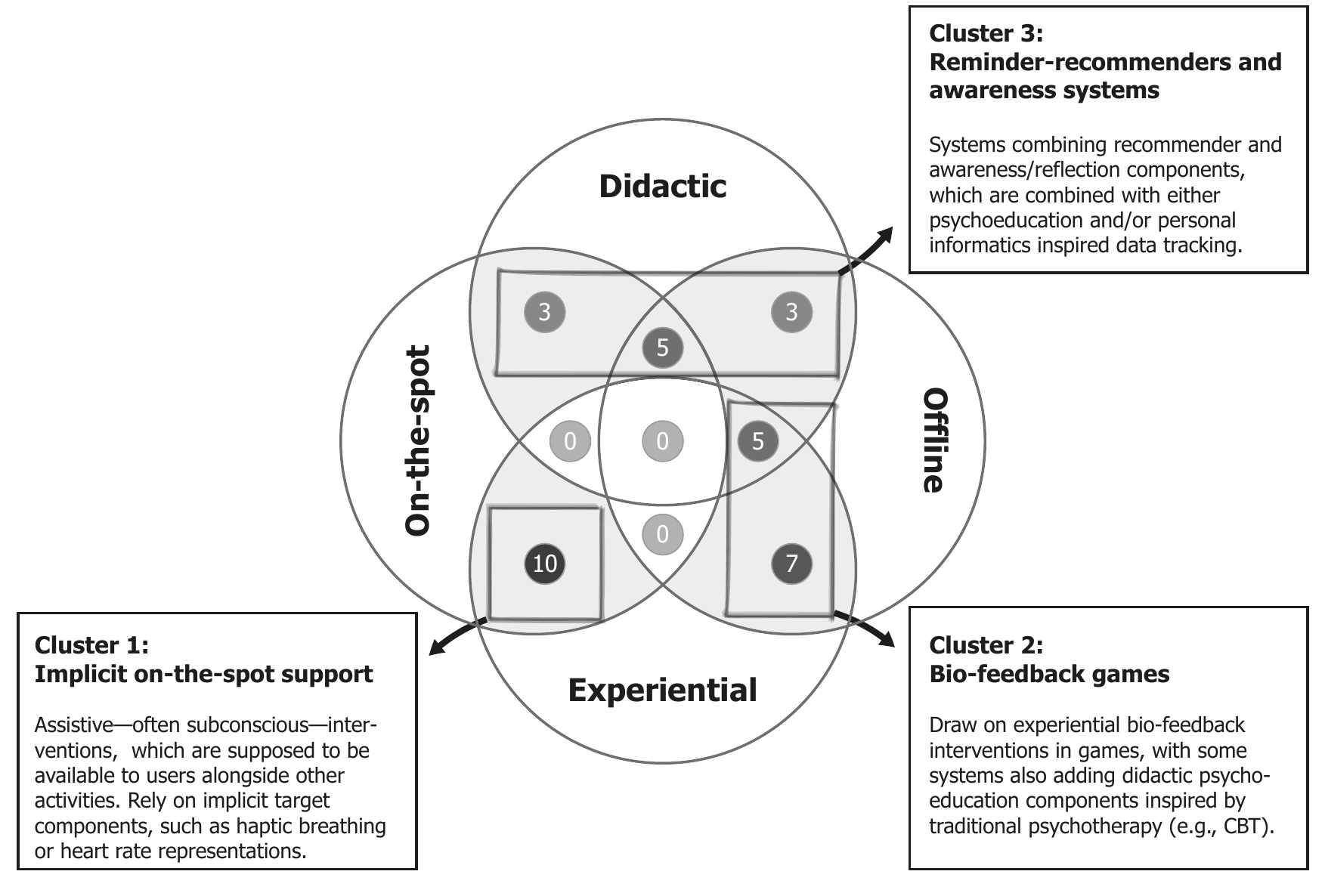} 
    \caption{Illustration of how the reviewed interventions are associated with the four delivery mechanisms -- didactic vs experiential; and offline vs on-the-spot. Numbers and location indicate the number of system using components within the respective intersection of dimensions (e.g., a system using components coded as didactic, on-the-spot, and offline would be placed in the appropriate intersection of the Venn diagram). The annotation shows the resulting clustering of research areas, as described below.}
    \label{fig:dimensionsDiagram}
\end{figure}

The previous section synthesised the \textit{design components} present in HCI work so far, with the aim to decompose existing work and provide us with the `dictionary' of techniques that have been utilised for each of the four possible delivery mechanisms (didactic, experiential, offline, on-the-spot).   
This section brings the focus back to the full interventions as presented in prior work to provide an overview of \textit{how the identified components are combined into systems} and identify any meaningful \textit{patterns or gaps of such combinations} across the current dataset. 

The key argument is that the existing work is, so far, clearly separated into three main research directions: \textit{implicit on-the-spot support}, \textit{bio-feedback games}, and \textit{reminder-recommender \& awareness systems}. 
These clusters correspond surprisingly neatly to the combination of delivery mechanisms used -- see Figure~\ref{fig:dimensionsDiagram}. 
In addition, the clusters are mostly independent of each other (draw on different background literature, intervention goals, and technological components), including using consistent combinations of design components within each cluster (i.e., components used in one cluster---e.g., implicit feedback or recommenders---are rarely utilised in other clusters)
Such lack of cross-cluster combinations suggests clear opportunities for future work as described in detail in Section~\ref{sec:cluster-summary} and Section~\ref{sec:research_agenda}: for example, there are several combinations of types of intervention components that have not been explored at all so far; as well as more intricate combinations of components that already exist. 


\subsubsection{Three clusters in the HCI work so far}

We outline each of the three clusters briefly below; the Appendix 2 then provides more details on the individual papers as well as the emotion regulation targets addressed in each of the clusters.

\paragraph{Cluster 1 -- implicit on-the-spot support (experiential + on-the-spot) -- n = 12} 
These systems explore assistive---often implicit---down regulation interventions,  which are supposed to be available to users alongside other activities. 
Systems in this category predominantly rely on experiential \textit{implicit target components} (n=9), drawing on a digital representation of a target state (heart rate, breath rate) -- cf., \cite{Balters2020, Paredes2018,Costa2019, Choi2020, Miri2020}. 
Three additional papers showcase the potential for using alternative components to deliver experiential/on-the-spot interventions, predominantly by moving toward more consciously enacted emotion regulation strategies through physical objects \cite{Liang2018,Slovak2018}, or background biofeedback visualisation \cite{Yu2018}.

\paragraph{Cluster 2 -- bio-feedback in interactive games (experiential + offline +- didactic) -- n = 12}

The papers in this cluster predominantly draw on utilising experiential bio-feedback interventions in games (n=10), with some systems \cite{Antle2018, Antle2019, Knox2011, Scholten2016} also adding didactic psychoeducation components inspired or directly drawing on traditional psychotherapy (e.g, CBT in-person session with a therapist combined with a bio-feedback game \cite{Knox2011}). 
Most focus is on offline components to \textit{both elicit emotion and scaffold the associated emotion regulation} \cite{Lobel2016, Mandryk2013, Parnandi2017, Parnandi2018, Wang2018, Zafar2017, Scholten2016} utilising a combination of interactive games (used to elicit mostly negative emotions such as stress or fear) and an experiential \textit{real-time biofeedback} loop component to provide feedback scaffolding ER training within the game space.
The remaining systems drew on some version of \textit{relaxation training} components \cite{Antle2018, Antle2019, Knox2011}, combining traditional biofeedback loops with simple game-like interactions (e.g., blowing on a virtual windmill).

\paragraph{Cluster 3 -- reminder-recommender \& awareness systems (didactic + offline || on-the-spot) -- n = 11}
The work in this cluster combines the existing work on \textit{reminder-recommender} and \textit{awareness and reflection} systems, potentially combined with either \textit{psycho-education} or \textit{visualising patterns over time}.
The papers are more diverse in terms of components used and their combinations in contrast to Clusters 1 \& 2, but this also comes with a less well established groundwork on the assumed theories of change: for example, only 2 out of the 11 papers \cite{Huang2015, Smyth2016} were coded as High on intervention model specificity -- cf., Appendix 1. 
The systems including \textit{aware\&reflect} \cite{Bakker2018, Huang2015, Kocielnik2013, Smyth2016, Wang2019} involved monitoring and tracking individuals’ emotions over time as a means to increase users' awareness of their emotional state(s); most of this work is akin to---and likely inspired by---HCI work on personal informatics systems (cf., \cite{Epstein2015}). 
In contrast, the systems relying on \textit{reminders or recommenders} provided the users with on-the-spot suggestions of specific ER strategies to use \cite{Diaz2018,Fage2019,Huang2015,Paredes2014,Pina2014,Smyth2016}. Conceptually, these systems often draw on behavioural change intervention systems, especially the Just-in-time-adaptive-intervention literature (see e.g., \cite{Nahum-Shani2018} for a recent review).
Only two systems have combined both aware\&reflect and reminder-recommender components \cite{Huang2015,Smyth2016}. 



\subsubsection{Cluster X: identifying missing design component combinations} 
\label{sec:cluster-summary}

As illustrated in Figure~\ref{fig:dimensionsDiagram}, there are three potential combinations of component types that had not been seen in the current dataset. Specifically, this involves interventions that would combine \textit{all} types of component (\textit{didactic, experiential, on-the-spot, offline}) as well as systems including a combination of \textit{experiential + on-the-spot} with either \textit{didactic} or \textit{offline} components. 
While the next section (Section~\ref{sec:research_agenda}) will describe how this and other research gaps highlighted by the proposed framework can drive and inspire future HCI work in more detail, we wanted to share a few observations already here, while the context of the work presented in the clusters above is still fresh in the reader's mind. 

On one hand, as researchers thinking about ER interventions in terms of intervention components, we find the lack of specific combinations within the existing intervention systems literature interesting and somewhat surprising: 
For example, we can see nothing that would logically prevent bio-feedback game interventions such as those described in Cluster 2 from including an additional on-the-spot recommender component, especially if the game-based learning is already complemented with didactic psychoeducation components (e.g., talking therapy). In fact, such recommender systems would logically complement---and perhaps could help resolve---the difficulties with transferring the learning from offline game experiences into on-the-spot applications. 
Similarly, there are no psychological reasons why a Cluster 1 intervention (e.g., a task-based breathing scaffolding device) could not be complemented with a form of aware\&reflect system, or a didactic psycho-education component that might promote mindful attention to breathing (which, arguably, could be then amplified on-the-spot by the haptic feedback already present in Cluster 1 systems). 

On the other hand, the lack of cross-pollination of techniques across clusters is less surprising if seen from the perspective of HCI researchers, given the different historical background that the existing interventions draw on, the relatively early stage of this research area in HCI, and often the resulting focus on investigating a particular interaction design technique as the research contribution (cf Appendix 1).
Perhaps unsurprisingly then, while the interventions within each cluster might be thorough in referencing other related work in that cluster, there are far fewer references to work \textit{across} clusters. For example, self-awareness or recommender systems may not appear immediately related to bio-feedback game interaction, especially if seen from the perspective of an HCI designer trying to deeply understand the interaction design characteristic that would make, say, a bio-feedback overlay on a game acceptable and effective in communicating the relevant psychological information. 

It is our hope that the proposed framework---and more broadly, this paper---can help highlight these less immediate connections across clusters and promote a deeper interconnection of work within HCI and (clinical) psychology collaborators, especially as the technology-enabled ER research area matures over the years. 
The next section goes on to outline the key gaps and opportunities we see, based on the analysis so far, as particularly interesting for the HCI and psychology communities to address in the next stages of the work. 




\section{Discussion and research agenda}
\label{sec:research_agenda}

Our purpose in writing this paper was to make sense of this rapidly growing and inherently interdisciplinary research area. 
The proposed framework is our way of communicating the understanding of what has been done so far (as synthesised in previous sections), but also a way of helping shape what comes next: we will show how the three parts of the framework (theory, strategy, practical) can be applied---separately or together---as lenses to identify gaps in existing work, inform new projects, and support ongoing synthesis of work across the range of disciplines and possible research threads. 

In the following subsections, we aim to \textit{model how we think about using the framework to generatively shape questions for the field}. 
We will start by briefly outlining what we see as the core remaining gaps when viewing the field from the perspective of focusing on (i)~the possible ER targets and psychological impact of our systems (theory \& practical); and (ii) what we know about the interaction design approaches for each of the specific intervention delivery mechanisms (strategy \& practical). 

We then discuss the cross-cutting considerations of how the framework can be applied more practically to guide the design of specific research projects, across all stages between traditional HCI focus (e.g., technological and interaction design innovation through exciting prototypes) and  psychology focus (large-scale efficacy trials on well-understood, robust, in-situ systems). 
%
%
%
We hope this reflection will be useful not just for individual researchers sharpening their research questions and approaches, but also to allow us---as a community---to have rich discussion of the field, gaps and opportunities, creating pathways towards a deeper collaboration among this rich set of disciplines. 

\subsection{Research gaps -- ER targets perspective} 
One way of identifying remaining gaps is through foregrounding considerations of the psychological targets our interventions aim to impact: i.e., the psychological processes necessary to developing ER skills and thinking about how these are---or are not---supported by \textit{any} of the interaction design components we see across the literature so far. 
%

\subsubsection{Lack of transfer support} 
Transfer is a crucial part of ER interventions. Enabling people to, e.g., better cope with daily setting \textit{without ongoing support} is crucial for long-term impact and sustainability of effective systems. Lack of transfer-enabling support in-situ is one of the key gaps in existing non-technological interventions (cf., Section~\ref{sec:step1-all}). However, as discussed in Section~\ref{sec:HCI_overview}, less than 15\% of interventions were specifically designed to reduce end-user dependence on the intervention system, i.e., to enable the end-user to develop and then use emotion regulation strategies on their own (\cite{Antle2019,Antle2018,Scholten2016,Zafar2017,Slovak2016}) -- and most of these relied on non-technological approaches to transfer support (e.g., \cite{Antle2019,Antle2018,Slovak2016}.  The remaining 31 interventions (implicitly) assumed that the developed systems would have to provide ongoing support for the intervention to be effective, and thus the effects would disappear if the system (such as a wearable feedback system) was removed.

We expect the questions around supporting transfer and skills development will be a highly fruitful avenue of research in the near future: it is a challenge which innovative tech-enabled intervention delivery mechanisms could be well fitted to address, but also one which has not been explored so far and, thus, a direction where likely many low-hanging opportunities exist.  
Within the framework model we see transfer as designing innovative connections between components sitting at the opposite ends of the \textit{wheres/whens} and \textit{hows} spectrums: for example, connecting what is learned in an offline component (e.g., bio-feedback game) with scaffolding on-the-spot application (e.g., reminders, on-going haptic feedback, or other new strategies) is one promising approach to addressing this crucial challenge. Similarly, envisioning approaches that combine didactic components (such as cognitive restructuring) with experiential approaches (such as on-the-spot experiential support) are likely to lead to increased transfer effects. 

\subsubsection{Uneven support for ER strategies} 
As outlined in Section 4, the support for individual ER strategies (situation selection, situation modification, attentional deployment, cognitive change, response modulation) is not evenly distributed: across all delivery mechanisms, most of the interactive components are targeting \textit{response modulation} strategies (although these are theoretically known to be less effective in the long term), with other well-established strategies such as \textit{cognitive reappraisal} or \textit{attentional deployment} virtually unsupported. This is particularly pertinent for on-the-spot and experiential components, and the reliance on simple reminders and/or non-technological module delivery in others. We also note that recent work suggests that digital technologies are already often used by participants for \textit{situation selection} as part of everyday use (cf., \cite{Wadley2020}), but these approaches are yet under-represented in the interventions we see in our study sample. 
A similar picture emerges with the stages of the ER process (identification, selection, implementation, monitoring), where most systems predominantly focus on supporting \textit{implementation} of specific strategies (i.e., mostly response modulation), and the remaining aspects remain  unsupported.

\subsubsection{Lack of efficacy data on (multi-)component systems}
\label{sec:RGap-ER-efficacy}
To date, there is a lack of data about efficacy of any of the four components that would provide evidence that these approaches lead to intended ER effects. This may be a result of the combination of several factors. Given the early stage of most HCI research to date, there is a lack of robust, hi-fidelity prototypes that can be deployed in controlled or RCT style studies. Only 9 papers (e.g. \cite{Antle2019, Fage2019, Costa2019, Knox2011} were classified as late stage in which systems were ready for widespread deployment. Second, of the 25 papers that did have a focus on intervention efficacy, few had outcome measures based on ability to select and implement (or monitor) specific ER targets (exceptions \cite{Fage2019, Antle2019, Scholten2016}). Other measures often included pre-post test ratings related to stress, anxiety, and depression (e.g. \cite{Knox2011, Scholten2016, Lobel2016, Antle2019}), performance on follow-up tasks (e.g. \cite{Zafar2017}) and some studies also logged and analyzed physiological data during sessions (e.g. \cite{Mandryk2013, Wang2018, Wells2012, Lobel2016}). Lastly, evaluation methodologies rarely included experimental designs, with only 4 field-based experimental studies \cite{Knox2011, Fage2019, Antle2019, Antle2018} and two RCTs \cite{Scholten2016, Wells2012}. Instead, the methodologies relied on observational (e.g. \cite{Mandryk2013, Lobel2016}) or various forms of comparative designs (e.g. \cite{Wang2018, Wang2018, Smyth2016}). This is a common challenge in HCI research (cf., \cite{Sanches2019}), drawing on the difficulties of resourcing a full development from inception to prototypes to robust, large scale deployments necessary for testing efficacy -- see also Section~\ref{sec:discussions_interdisciplinaryCommunity} (and cf., \cite{Coyle2009} on analogous arguments in technology-enabled mental health interventions more broadly). 

What was also noticeably missing were evaluations that collected data during the course of a study to evaluate \textit{process measures} (exception \cite{Smyth2016}). While many early stage studies explore interactional processes (e.g. interpretation of different modalities of data representation), few looked at any type of interactional processes that might lead to ER effects, such as engagement and perceived psychological impact of the intervention \cite{Slovak2018} or learning trajectories \cite{Antle2018}. One of the benefits of including process measures is that they can be used to assess the validity of the theory of change and related logic model in early deployments (e.g., see optimisation studies used in behaviour change research \cite{Collins2011,Boruvka2015}).  Examination of both psychologically valid outcome and process measures related to learning, mastering, and deploying ER strategies in real world situations is another promising area for future work.


\subsection{Research gaps -- intervention delivery mechanisms \& their implementation perspective} 
Another approach to identifying promising future work directions is in foregrounding the design considerations around plausible intervention delivery mechanisms: i.e., what do we not yet know about the possibly innovative ways in which ER intervention components could be delivered to our populations? What are the novel approaches to supporting the \textit{hows}, and the \textit{wheres/whens}?

In what follows, we again highlight what we see as the most interesting questions, separated according to the delivery mechanism types as identified in the strategy part of the framework (cf., Section~\ref{sec:conceptualDim_articulation}). 
Each of the paragraphs starts with a reminder of the definition for the respective components, and follows up with the corresponding gaps in existing literature. 

\subsubsection{Didactic} \noindent
These components focus on the delivery of information and supporting its conscious application, even if it is not immediately done within the targeted social-emotional context.

\begin{enumerate}
	\item Most of the systems focus on information delivery, but far \textit{fewer technology-enabled approaches are available for supporting users to put these into practice}, whether that is within real-world situations or within out-of-context practice examples. For example, can technology provide step-by-step (but still meaningful) scaffolding to practice newly trained skills? What would it mean to re-design the existing reflection support to provide more direct intervention support, moving beyond simple self-awareness or pattern detection (e.g., emerging EMA-as-intervention approaches \cite{Balaskas2021})?
	\item We lack research on \textit{innovative approaches to information delivery that would embed didactic learning into participants' lives} (rather than relying on traditional module-based delivery or simple reminders). For example, how might technology help decompose/personalise the existing cognitive modules into bite-sized pieces, e.g., for the learning to be accessed based on immediate need? How might existing/new HCI techniques draw on traditional role-play techniques to amplify the learning of ER skills  (e.g.  story-based learning, embodied games, engaging fictional worlds \& narrative)? 
	\end{enumerate}

\subsubsection{Experiential} \noindent These components focus on mediating the first-hand experience of applying a specific emotion regulation strategy. In other words, such components provide scaffolding in support of the performative aspects of skills development. 

\begin{enumerate}
	\item We have yet to see research on how \emph{performative aspects of well-known adaptive ER strategies} (such as cognitive change, attentional deployment or even situation modification) \textit{could be experientially supported through interactive technologies}. For example, how might cognitive reappraisal \textit{practice} be supported through a digital system (e.g., through scaffolded narrative experiences, ongoing reflection, or other ideas?). What would be alternative interaction design mechanisms to provide experiential support beyond bio-feedback or on-body haptics? 
	\item As a related aspect, we so far lack research on \textit{how technology could directly scaffold and/or guide users through the emotion regulation experience}. For example, how might we apply known (or envision new) interaction design mechanisms to \textit{guide} users through emotional experience trajectories \cite{Benford2009, Tennent2021}, rather than ‘just’ giving feedback on their emotional state? How might these opportunities be supported through emerging interaction techniques such as embodied games \cite{Isbister2018, isbister2016games}, digital arts \cite{Benford2009}, or mixed reality devices?
\end{enumerate}

%



\subsubsection{Offline}
The main characteristic of offline intervention components is that they aim to support development of emotion regulation strategies in deliberate training contexts rather than in daily life. The trained skills are then expected to be transferred into everyday situations where strong emotions naturally arise.
\begin{enumerate}
	\item Existing work \textit{draws on a very limited range of interactive training contexts}: so far, the systems are limited to either (stress-inducing) video games, or relaxation support. For example, how might we utilise technologies such as VR, tangible devices, or interactive arts to facilitate emotion regulation training loops in a broader range of emotional/social contexts (including attention/cognitive reappraisal techniques)? What would it mean to draw on multiplayer experiences to enable new ways of experiencing/scaffolding emotion regulation techniques, whether that is in a game, virtual/mixed reality, or as an asynchronous interaction? 
	\item Similarly to much of behaviour change literature, \textit{existing systems lack innovative techniques to scaffold effective psychoeducation}, beyond the in-person techniques that are already well-established in clinical work. For example, how might offline design components such as bio-feedback be effectively coupled with (interactive) psychoeducation approaches? How could these be transformed to directly guide support user's sense making and learning from data collected in-situ (as such insights are often at core of psychoeducation approaches)? And finally, how might such systems be designed to specifically support training of individual stages of the ER process that have been so far under-represented in the current work (e.g., identification or monitoring)? 
\end{enumerate}

\subsubsection{On-the-spot}
These components aim to train emotion regulation strategies during naturally occurring emotional situations (e.g. stressful events in daily life). Therefore, the skills practised are immediately applied to the specific emotional situation, as mediated by the technological or in-person intervention; and—what is crucial—no further transfer of skills is expected or needed.

\begin{enumerate}
	\item  In contrast to the other three delivery mechanisms, \textit{we see a large range of potential techniques used (user/system/situation-specific; ongoing/discrete) but in effect, none of the existing interaction designs is explored in depth}. How might we better understand the benefits/costs of individual approaches in terms of learning impacts (e.g., when is it better to rely on user- vs system-initiated components, and when is a combination required?) 
	\item At the same time, on-the-spot components are \textit{so far relatively limited in the types of specific technological mechanisms used and emotional processes supported}: these are predominantly either simple reminders (`remember to reappraise your emotions') or some form of (on-body) haptic feedback. What other forms of on-the-spot delivery are feasible? Especially, how could we use interactive technologies to embed support for ER skill \textit{application} into moments of heightened emotional experiences in everyday life (e.g., similarly to the attention/response modulation impact of the Smart toys research stream \cite{Slovak2018,Theofanopoulou2019})? How might new systems utilise social interaction (for better/worse) to support emotion regulation practice, considering that most of our day-to-day interaction is in dyadic/group systems (e.g., families, households, friendship groups environments)?
	\end{enumerate}



\subsection{Guiding the interdisciplinary community development}
\label{sec:discussions_interdisciplinaryCommunity}


Finally, we discuss how the framework can support the emerging research community around technology-enabled emotion regulation in thinking about the field. 

We argue that---within such an inherently interdisciplinary space---it is important to support researchers and practitioners across the full scale from \textit{mostly HCI contributions} (e.g., exciting technology innovation with potential downstream ER impacts) to mostly \textit{psychological contributions} (e.g., measuring psychological impact of robust intervention systems tested in large scale deployments)\footnote{See e.g., \cite{Coyle2009} for an analogous process in the context of online-CBT research in the last two decades!}. 

As an example, we outline one possible way of structuring our thinking of such progression below, recognising the importance of research engagement at all levels of interest, resources, and research goals---with the framework providing a common language and conceptual structure across a range of technological/design readiness. 
We have added prospective HCI conference venues to further illustrate the type of work one might imagine being produced at each of the levels.

\begin{enumerate}[label=(L{{\arabic*}})]
	\item \textbf{theory-informed intervention component development} \\ 
	(UIST, UbiComp, DIS, CHI)\\ 
	\textit{i.e., focus on technology/ design innovation and exploration, but with a framework helping to infuse psychology theory into HCI design of potential intervention delivery mechanisms} \smallskip
	
	\item \textbf{focused investigation of supporting specific ER strategies}\\ 
	(DIS, CHI) \\ 
	\textit{i.e., focus on re-envisioning/validating mechanism implementations from prior level to understand their impact on ER targets, without necessarily developing a full intervention}\smallskip

	\item \textbf{combining existing design components into possible real-world interventions} \\
	(CHI, CSCW, JMIR) \\ 
	\textit{i.e., utilising successful implementations of delivery mechanisms in new ways, with the focus on understanding appropriation and potential for real-world impact} \smallskip

	\item \textbf{understanding for whom, where, and how proposed interventions work} \\ 
	(CHI, JMIR, main-stream psychology journals such as Lancet Digital Health) \\ 
	\textit{i.e., focus on efficacy and effectiveness data, including large-scale randomised trials and policy implications}\smallskip

\end{enumerate}

In each of these levels, the framework can orient HCI designers to the key design choices even if they are not psychology experts; and in the later stages, can help orient psychology researchers to the capabilities of technologies without requiring substantial interaction design knowledge.

In particular, we see the framework's separation of the `whats', `whens/wheres', and `hows' as guiding HCI researchers in:
(i) selecting appropriate ER targets and outcome measures as relevant for the granularity level (i.e., `whats'); (ii) building on each other's work as well as positioning it within the broader community (i.e., within the `whens' and `hows'); (iii) inspiring innovation through facilitating re-use and appropriation of interaction design components across contexts into effective interventions; and finally, (iv) progressively facilitating the transition of promising research prototypes from HCI into the hands of the psychology community, including large-scale real-world intervention deployments. 

In addition, the modularity inherent in the framework---e.g., thinking about technology components as implementing particular delivery mechanisms as part of the strategic whole---could promote better findings and transferability between what would be seen as the focus of different communities (e.g., those publishing in UIST vs JMIR).  

%



\section{Conclusions}

The main aim of this work is to help combine innovative HCI intervention approaches with substantial knowledge about the fundamental emotion regulation models from psychology. 
The proposed framework is then a way of synthesising the key aspects: encapsulating the psychology fundamentals (theory component, Section 2); articulating plausible technology-enabled delivery mechanisms that could be supported by innovative technologies (strategic component, Section 3), as well as the understanding of what has been done so far within HCI (practical component, Section 4). Finally, Section 5 highlights how such considerations can also help shape future research at the intersection of the HCI and psychology fields. 

Overall, it is our hope that the HCI research community can draw on the three parts of the framework---theory, strategic, practical---and the associated research agenda to support research at differing scale of granularity: from helping guide the overall research field as well as individual research programmes, to supporting specific design projects. 
On the \textit{research field level}, the framework aims to provide a common language and conceptual structure to connect research work across a range of technological/design readiness; and to enable rich discussions about the key research questions we---as an interdisciplinary community---should aim to address. 

On the \textit{individual research programme level}, we expect that individual researchers might most benefit from the practical framework component to help identify any remaining gaps in existing work, while positioning their work in the broader field with the help of the theory and strategic components. 
	For example, a Principal investigator (PI) identifies the relative lack of technology-enabled support for cognitive reappraisal (e.g., in contrast to response modulation), and is interested in understanding whether/how such interventions could be developed in VR (which is also missing from existing literature and the PI has expertise in).

\enlargethispage{\baselineskip}

Finally, on the \textit{specific study level}, the framework can both help sharpen the research question (in terms of specific gaps), while also enabling the PI to re-use and appropriate helpful techniques from across the fields. 
	For example, to continue with the cognitive reappraisal in VR narrative from above, a PI could use the framework as follows:
\begin{itemize}
	\item[-] The \textbf{theory component} guides the team in picking the respective ER target (e.g., specific cognitive reappraisal techniques to train) and outcome measures from previous experimental work in psychology. 
	\item[-] The \textbf{strategic component} helps systematise design decisions around the plausible delivery mechanisms (e.g., deciding to include a combination of offline, experiential, and didactic components). 
	\item[-] The \textbf{practical component} then helps identify specific design choices: for example, the PI might choose to include a \textit{didactic} psychoeducation component (inspired by prior DBT intervention) as an entry into the newly developed \textit{offline} VR experience (core contribution of the project), with \textit{experiential} support through biofeedback (re-using some biofeedback mechanisms from a game-based prior work) and emotional narrative experiences (also a core contribution, by re-purposing vignettes from therapy into immersive VR experience). 
\end{itemize} 

It is work across all these levels that is needed to help turn the current emerging interdisciplinary interest into a coherent body of research with the potential for delivering substantial real-world impacts on people's lives; and we hope this paper can invite, motivate, and support researchers from across HCI and psychology to join in on this journey. 


\bibliographystyle{ACM-Reference-Format}
\bibliography{references-Petr-Mendeley,review,refs-beyond-review,AlissaToCHIExpt}


\vfill~ \pagebreak
\section*{Appendix 1 -- Scoping review of existing HCI systems/intervention targeting ER}



In this section we provide a synthesis of research done in HCI related to the develop of technologically-based interventions that support learning, development, and practice of emotion regulation with adults and/or children. Overall, the main focus was to identify and code most if not all existing HCI work in terms of the psychological mechanisms behind the interventions (where possible), in addition to the more commonly addressed aspects (e.g., type of technology used). 

Specifically, we used an inductive and iterative approach to thematic analysis to differentiate existing systems along a priori selected six dimensions:  (1) type of technology; (2) stage of research to, e.g., distinguish early HCI explorations of materials from more developed systems; (3) topical focus of HCI research; (4) intervention mechanism underpinning the designed system; (5) level of intervention mechanism specificity; and (6) intended use-case (skills development vs continuous support). These dimensions have been derived from prior research and similar HCI reviews work, as well as the theoretical grounding from psychology (i.e., focus on theory of change, and learning vs on-going support). 

Following from the theoretical review in Section 2, we have also coded (7) the extent to which each system relied on in-the-moment (on-the-spot) vs out-of-context (offline) training; and cognitive (didactic) vs experiential (experiential) learning; these results are addressed in detail in Section 4.

\subsection{Methodology}
Specifically, we chose a scoping review methodology to develop a detailed synthesis of the type and range of studies available on this emerging topic in digital mental health \cite{Arksey2005_Scoping}: such approach is appropriate to identify the types of evidence available for a topic (which may be split across several fields), to examine how research has been conducted, and to identity knowledge gaps, which is in alignment with our aims.
A scoping review is conducted using a systematic procedure, which can be replicated and often aims identify knowledge gaps \cite{Munn2018_ScopingReview}. The full coded database of papers as well as the R code used for analysis will be made available through the Open Science Framework. 

In the context of the this paper, a key component of our scoping review was the rationale for the inclusion of six core dimensions, followed by iterative inductive thematic analysis \cite{Tesch1990} to iteratively cluster and (re) define categories and codes within each dimension, creating an analytical framework which we then apply to the entire set of selected papers. Our results provide a broad yet in-depth descriptions of research in this space.
%
%
This approach enabled us to synthesise high level themes related to the kinds of research that has been done related to interactive technology development for learning, developing, or practising ER, with a specific focus on research with an end goal of creating ER interventions. The results of our scoping review provide a synthesis of research to date, serving as a foundation that enables us to highlight gaps or opportunities for HCI, which we then discuss in the remainder of this paper. 

\begin{figure}
    \centering
    \includegraphics[width=0.8\textwidth]{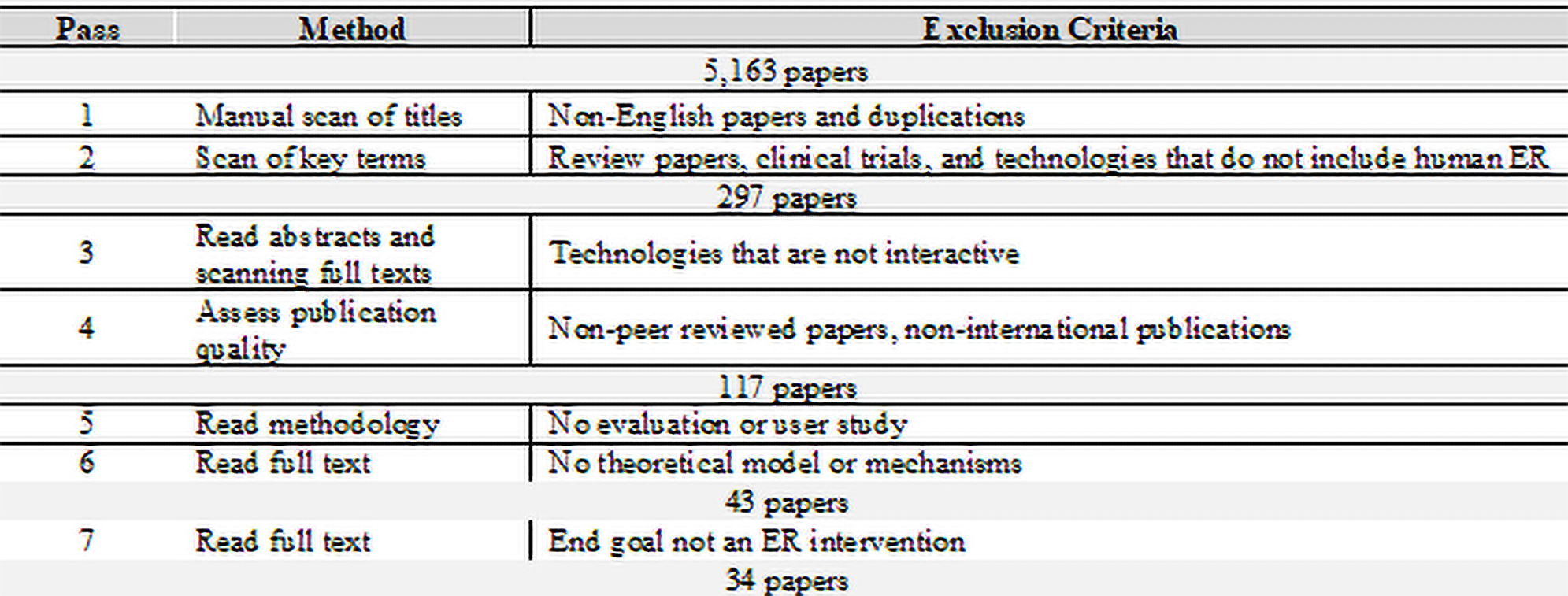}
    \caption{Literature review exclusion criteria Placeholder}     \label{fig:lit-rev-excluded}
\end{figure}

\paragraph{Selection and Filtering:}\quad 
We conducted a literature review that targeted full papers and notes as well as high quality works-in-progress that were published in top venues in HCI. We targeted the ACM digital library, IEEE databases and used Google Scholar to search for papers published between January 1 2009 and December 8 2021. Our keyword search criteria were “interactive technology” and one or more of “emotion regulation” or “self-regulation” or “stress” or “stress-regulation”. The search resulted in 5,574 papers.
We then conducted two passes of filtering (see Fig~\ref{fig:lit-rev-excluded}) based on title and keywords,  excluding papers that were non-English, duplicates, review papers and/or did not address human emotion-regulation, leaving 333 papers. 

During the third and fourth passes of filtering, we scanned abstracts and/or full texts and excluded papers that did not involve an interactive technology and reviewed publication venues, removing non-peer reviewed papers and papers that were published in non-international publications, leaving 130 papers. In the fifth and sixth passes of filtering we reviewed the full text and excluded papers that did not include an evaluation or a user study of any kind or did not include the identification of one or more intervention mechanisms, leaving 52 papers. 

Finally, we conducted a seventh pass in which we read the full text and excluded papers where the end-goal of the research was not an emotion regulation intervention (resulting in 36 papers), which was necessary for the follow-up analysis described in Section~\ref{sec:framework}. This is the dataset that we report on in this section.  

\paragraph{Inductive Analysis Process}\quad   
Our analysis process followed an inductive interpretive analysis process, similar to open coding in grounded theory \cite{Corbin1990}, since we did not have an established analytical framework for coding the papers that were selected in our scoping review \cite{Arksey2005_Scoping}. In this process, we iteratively identified  themes or dimensions, which describe the research papers from different perspectives. For each dimension we iteratively and comparatively developed and defined categories, resulting in the development of a codebook, which we then used to code the entire dataset. 
We used our research goals and focus on ER interventions to structure this inductive and iterative process, beginning with  the two researchers (first and last authors) working together to identity the six dimensions that of the analysis framework, as described below.  The dimensions were derived after reading through a quarter of the papers and skimming many others. We worked individually and then together to derive the following six dimensions (see \ref{table:dimensions} for an overview):  

We included the \textit{(1) type of technology} dimension because it enabled us to code and quantify the kinds of hardware (e.g. input, output, platform) currently being used to create ER interventions. 
We also examined where each research study fell on a typical research \textit{(2) stage in research lifecycle} (i.e. early, mid, late stage), which helped us characterize the maturity of each study and better understand current state of the field. 
Third, we described the underlying \textit{(3) HCI research focus} by identifying clusters of HCI research that shared similar foci (e.g. input and output modality research, system co-design research, intervention efficacy research), representing the different theoretical foundations and the range of methodological and interdisciplinary traditions found in HCI research as applied to ER intervention context.  
To explicitly capture the psychological components of interventions, we then coded for \textit{(4) the intervention mechanism} -- that is the psychological 'theory of change' through which the intervention is assumed to affect emotion regulation (e.g. Reminders, Biofeedback);   
(5) the \textit{level of specificity of that mechanism} (i.e. high, medium, low); and 
the intended (6) \textit{ongoing vs skill-development support}, i.e., whether the intervention was designed to be used indefinitely (and effects would disappear if taken away), or aimed as a temporary scaffolding for skills development.  The last dimension is particularly important given the ramifications for the long term effectiveness of interventions, as well as the learning theories underpinning the intervention designs.


Once dimensions were identified, the dataset analysis phase consisted of two of the researchers working inductively to iteratively to develop initial categories  and  coding rules  for each of the six dimensions. This was done during individual and shared analysis sessions, each focusing on a subset of the dimensions (e.g. (3) HCI focus and (4) intervention mechanism), over a small subset of papers. Once categories were defined, the researchers then individually reviewed and coded about 25\% of the papers across all six dimensions, together reconciling, revising and finalizing  descriptions of categories and coding rules for  each of the categories. This codebook was finalised after we had read and analyzed about half the papers, at which we had well-defined dimension categories and stable coding schemes for each dimension. 

As such, the resulting categories within each of the dimensions have predominantly emerged from the data and are thus presented together with the results overview below.

\begin{figure}
    \centering
    \begin{subfigure}[b]{0.45 \textwidth}
        \centering 
        \includegraphics[width = \textwidth]{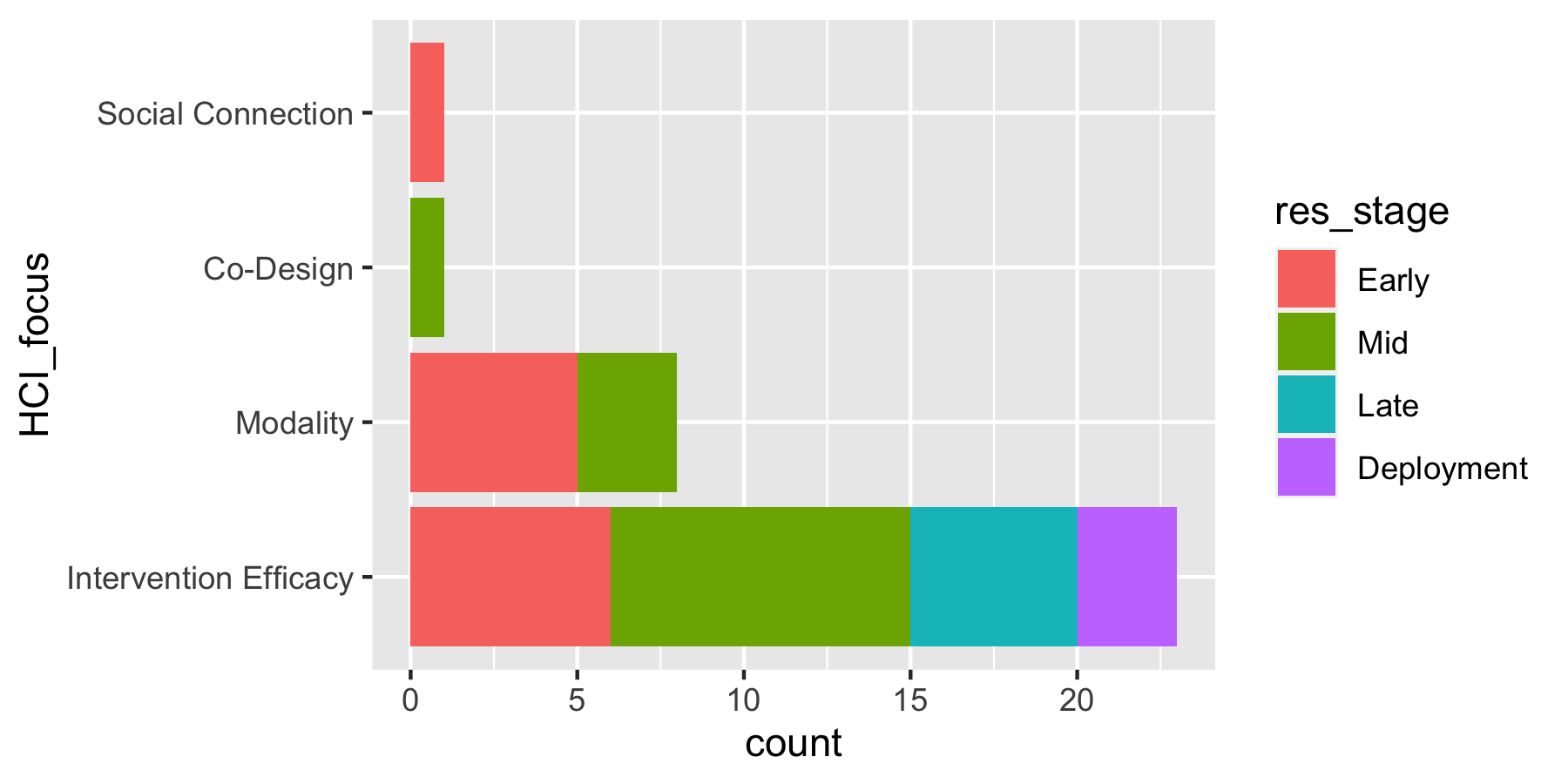}
        \caption{HCI focus by stage}
    \end{subfigure}
    \begin{subfigure}[b]{0.45 \textwidth}
        \centering 
        \includegraphics[width = \textwidth]{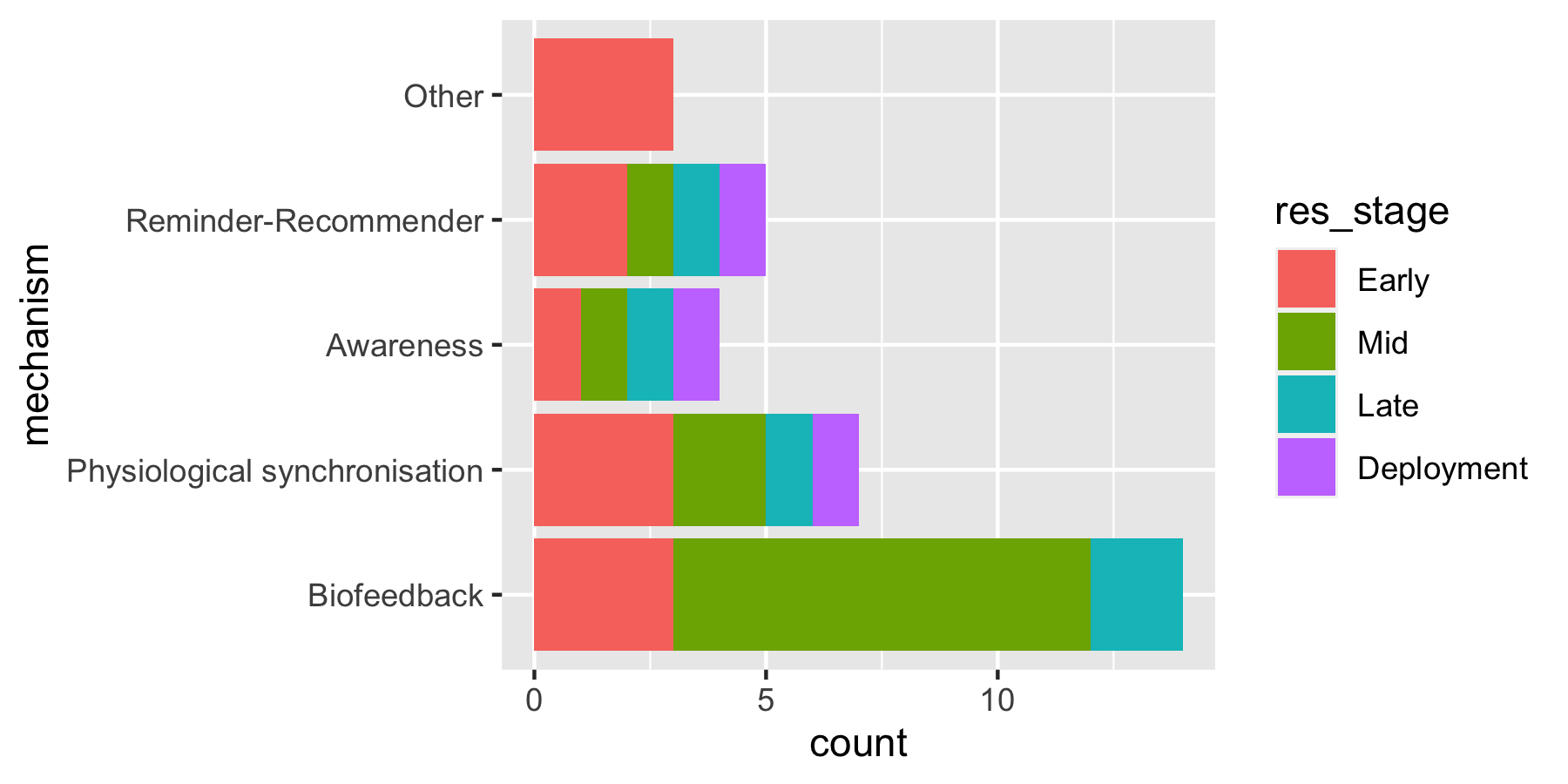}
        \caption{Mechanisms by stage}
    \end{subfigure}
    \caption{Overview of the interaction between stage-of-research and mechanisms and HCI focus codes.}
    \label{fig:stage-distribution}
\end{figure}

\subsection{Results}
Overall the papers are comprised of 36 peer-reviewed papers, split 63.9\% ACM, and 19.4\% IEEE databases, with the remaining 16.7\%  available through Google Scholar. 
%
%
In what follows, we first outline the results for each of the dimensions separately, and then highlight some of the arising connections. Each section starts with an overview of how the categories within a dimension  are represented in the dataset.

\subsubsection{Type of technology}
We classified papers by the type of hardware technologies used to implement prototypes or systems. When looking at the dataset, the most common platforms were mobile devices (16) and desktop or laptop computers (11), with several other devices such as  smart watches (4), and bespoke embedded computing prototypes (10). Seven papers described systems that were an assembly involving custom hardware (i.e. system was custom-made assembly of devices and electronic parts), the remainder of papers involved systems created from combining one or more commercially available devices. The sensor(s)—connected to the main platform for systems—included a variety of biosensors (e.g. respiratory, HR, EMD, EEG, GSR sensors) and non-screen based output devices(s) which included VR headsets, tangible fidget, lamp display, haptic seat pad).

\begin{figure}
    \centering
    \begin{subfigure}[b]{0.45 \textwidth}
        \centering 
        \includegraphics[width = \textwidth]{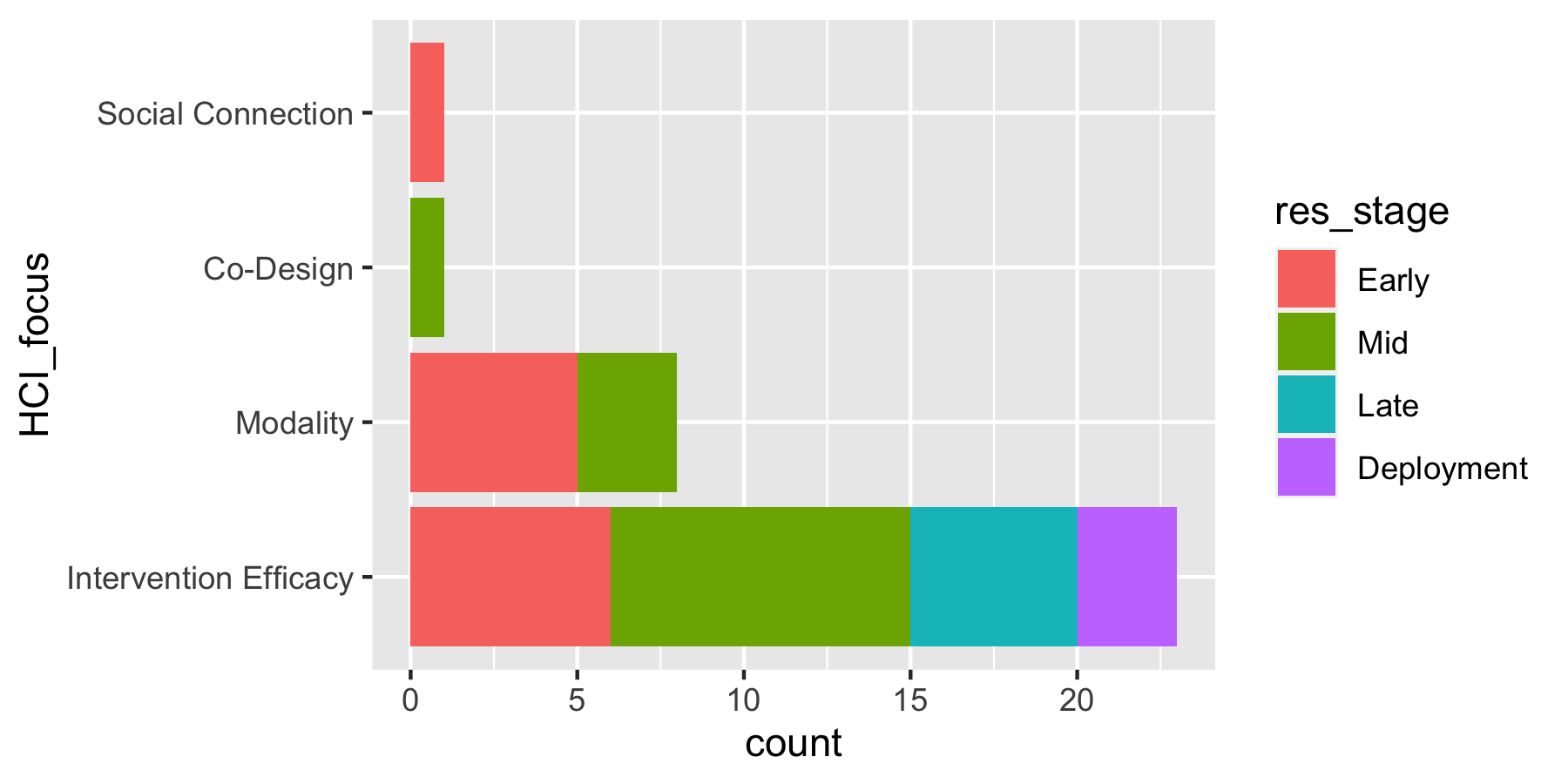}
        \caption{HCI focus by stage}
    \end{subfigure}
    \begin{subfigure}[b]{0.45 \textwidth}
        \centering 
        \includegraphics[width = \textwidth]{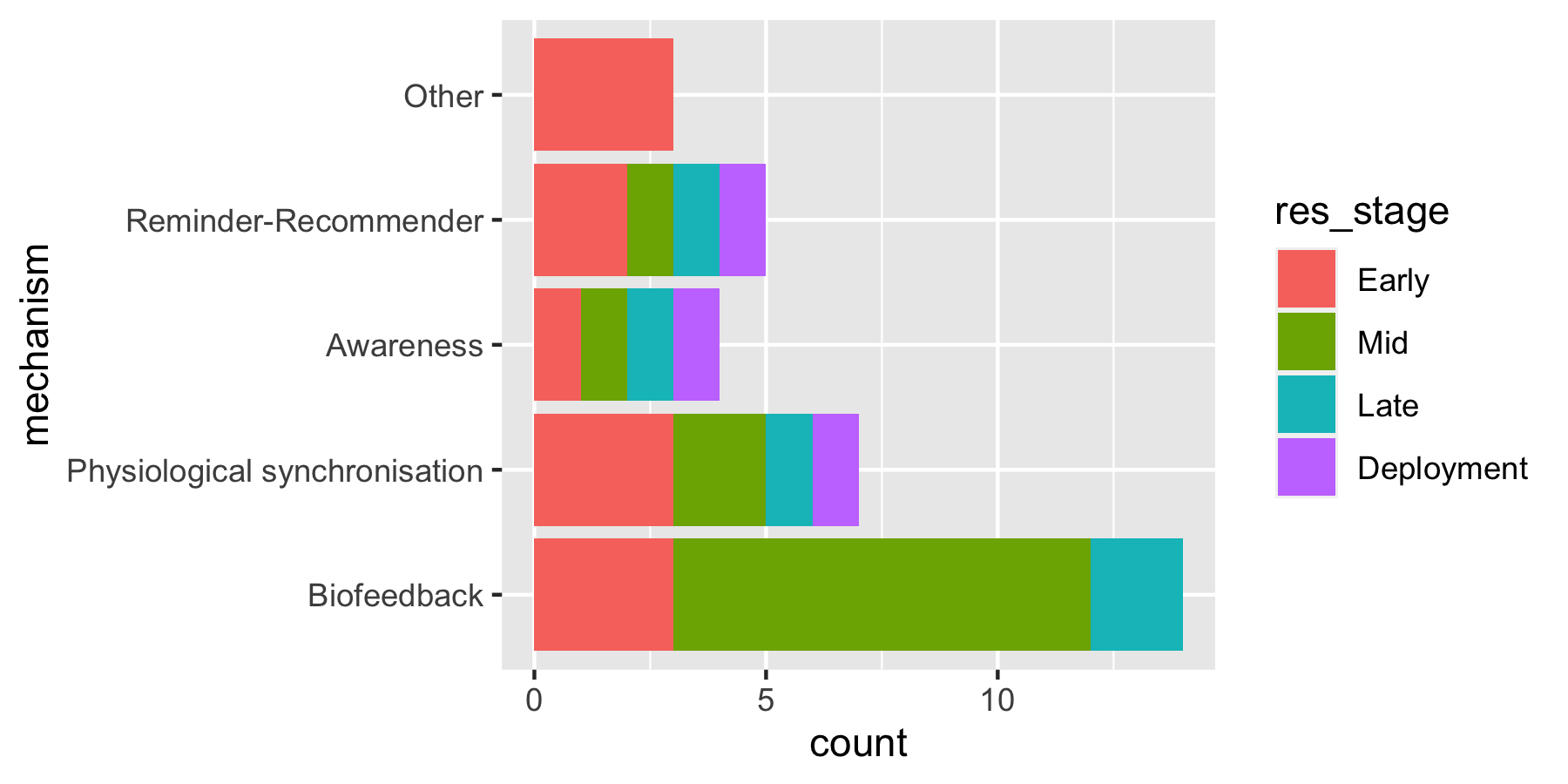}
        \caption{Mechanisms by stage}
    \end{subfigure}
    \caption{Overview of the interaction between stage-of-research and mechanisms and HCI focus codes.}
    \label{fig:stage-distribution}
\end{figure}

\subsubsection{Stage of research}
\label{sec:appendix1-research-stage}
HCI investigations may take place over a broad 'lifecycle' of design research, spanning from early design explorations to large-scale in-the-wild studies. We classified papers based on where the described \textit{interventions} were on this research trajectory. To simplify matters, we coded the papers as early, mid, or late stage. This qualitative assessment incorporated a range of factors, spanning the research questions addressed by paper authors (e.g., examining viability of a new design concept or form factor, versus efficacy evaluation of robust systems); the purpose and methodology of the evaluation (formative vs summative study designs); as well as the robustness of the examined prototype (e.g., testing a component of a future-envisioned system in-lab, versus in-situ deployments of robust full system implementation). 

Specifically, the 13 papers we coded as \textit{early stage research} focused on the design of prototypes and formative evaluations. The 14 papers coded as \textit{mid-stage research} focused on iterative design and evaluation of more complex and/or robust prototypes with formative or summative evaluations. Finally, the 9 papers coded \textit{late stage research} involved hi-fidelity, robust research systems that were ready for wide-spread deployment and summative evaluations in lab or field studies.  The Figure~\ref{fig:stage-distribution} shows how these papers are divided across the two key coding categories described below: HCI research focus, and the psychological mechanism.


\subsubsection{HCI research focus} %
\label{sec:appendix1-research-focus}
The papers in the data set differed markedly in their theoretical grounding (from post-positivist to critical design) and the contributions they attempt to make (e.g., from verifying intervention efficacy to introducing innovative design ideas). In identifying the `HCI research foci', our aim was to characterise the commonly used patterns in the existing body of work, similarly to how papers might be clustered together at an ACM CHI session. The resulting groupings and example papers are as follows: 

\begin{itemize}[label=-, leftmargin = 8pt, itemsep = 4pt, topsep= 4pt, parsep=4pt]

    \item \textit{Intervention Efficacy Research (n=25)} involves research in which the focus and motivation are on evaluating the effectiveness of a technology-mediated intervention in terms of improving learning, developing or practicing ER in a variety of contexts (e.g. lab, home, everyday activities). Of these most papers were mid to late stage research, with the exceptions tended to focus on early stage feasibility studies (e.g. \cite{Slovak2018, Diaz2018, Kocielnik2013}. The evaluations were split across all intervention mechanisms described below in section~\ref{subsubsec:intervention_mechanisms}, including Awareness (5), Physiological Synchronization (3), Reminder and Recommendations (5), Biofeedback (10) and Other (2): 
    
        In studies of systems based on mechanisms of Awareness, Physiological Synchronization and R\&R, research designs tended towards mixed measures observational studies. This was also true for one of the papers categorised as `Other’ \cite{Slovak2018}, which was based on in-situ haptic calming. While the results of many of these studies were largely positive, the measures used typically focused on end-users’ ability to use and understand the systems involved in interventions, rather than evaluating changes in skills development around ER. Most evaluations resulted in findings related to the need to change aspects of the design of the intervention. 
    
        Biofeedback-based interventions tended to be evaluated at a higher level of rigor. Those at mid or late-stage were either observational studies (e.g. \cite{Mandryk2013, Lobel2016}) or various forms of comparative (e.g. \cite{Wang2018, Wang2018, Smyth2016}) or controlled experiments (e.g. \cite{Knox2011, Zafar2017, Antle2019}). There were two RCTs \cite{Scholten2016, Wells2012}. Measures often included pre-post test ratings related to stress, anxiety, and depression (e.g. \cite{Knox2011, Scholten2016, Lobel2016, Antle2019}), performance on follow-up tasks (e.g. \cite{Zafar2017}) and some studies also logged and analyzed physiological data during sessions (e.g. \cite{Mandryk2013, Wang2018, Wells2012, Lobel2016}). In many of these studies participants were often children with ER challenges (e.g. ADHD, anxiety , fetal alcohol syndrome). In general, results were often mixed with positive evidence related to showing a direct impact of the intervention on participant’s ability to regulate stress and/or anxiety during the intervention (for an exception see \cite{Scholten2016}). However, lack of controls, short duration of interventions and/or lack of direct measures of ER skills development measures outside of the intervention limit validity and generalizability. Only two studies measured transfer of ER skills into everyday life or administered a follow-up test to determine maintenance of effects \cite{Antle2018,Antle2019}. 
    
    \item \textit{Modality Research (n=9)} involves work in which the focus and motivation are on evaluating the effectiveness of a technology-mediated intervention in terms of improving learning, developing or practicing ER in a variety of contexts (e.g. lab, home, everyday activities). Modality papers were entirely early to mid-stage research. Out of these, several papers focused on comparing different forms of outputs to support physiological synchronization, including a comparison of placement for haptic actuators representing heart rate \cite{Miri2020}, and comparisons of different representations of breath (e.g., haptic versus voice \cite{Paredes2018}; tactile versus visual and auditory \cite{Choi2020}; pacing \cite{Moraveji2011}). Several other papers investigated modalities utilized in biofeedback systems, including the use of visual metaphoric representations of stress over time \cite{Yu2018}, the interplay of breath and heart rate with tangible and light outputs \cite{Liang2018}, a comparison of different physiological inputs \cite{Parnandi2017} and exploring the timing and duration of feedback \cite{Parnandi2018}. These types of explorations are necessary to understand the impact of design decisions on user experience as part of the iterative development of systems that can be used interventions.

    \item \textit{Co-Design and Social Research (n=2)} was defined as either (i) research in which the investigation focuses on and is motivated by better understanding how to design ER prototypes and interventions through co-designing systems and/or exploring end-user customization of prototype emotion regulation systems (\cite{Parades2014}) ; or (ii) as investigating how to design to support social interaction as a key element of emotion regulation interventions (\cite{Slovak2016}). Suprisingly, there was only one example of these otherwise common directions in HCI research in this dataset. 
    
    However, some of the papers in different HCI foci areas also mentioned the importance of co-design and/or personalization, for example of design choices relative to placement of haptic actuators \cite{Miri2020}). Similarly, other papers mention the importance of social factors to intervention efficacy, as a secondary consideration. For example, the importance of facilitator rapport is cited as a key factor leading to positive results in a biofeedback intervention for children \cite{Antle2019} and teacher created content for interventions \cite{Diaz2018}. These and other studies suggest the benefits and importance of considering the social landscape of ER interventions, however, this area is currently under-represented in HCI as a focus of research.

\end{itemize}

\begin{figure}
    \centering
    \begin{subfigure}[b]{0.45 \textwidth}
        \centering 
        \includegraphics[width = \textwidth]{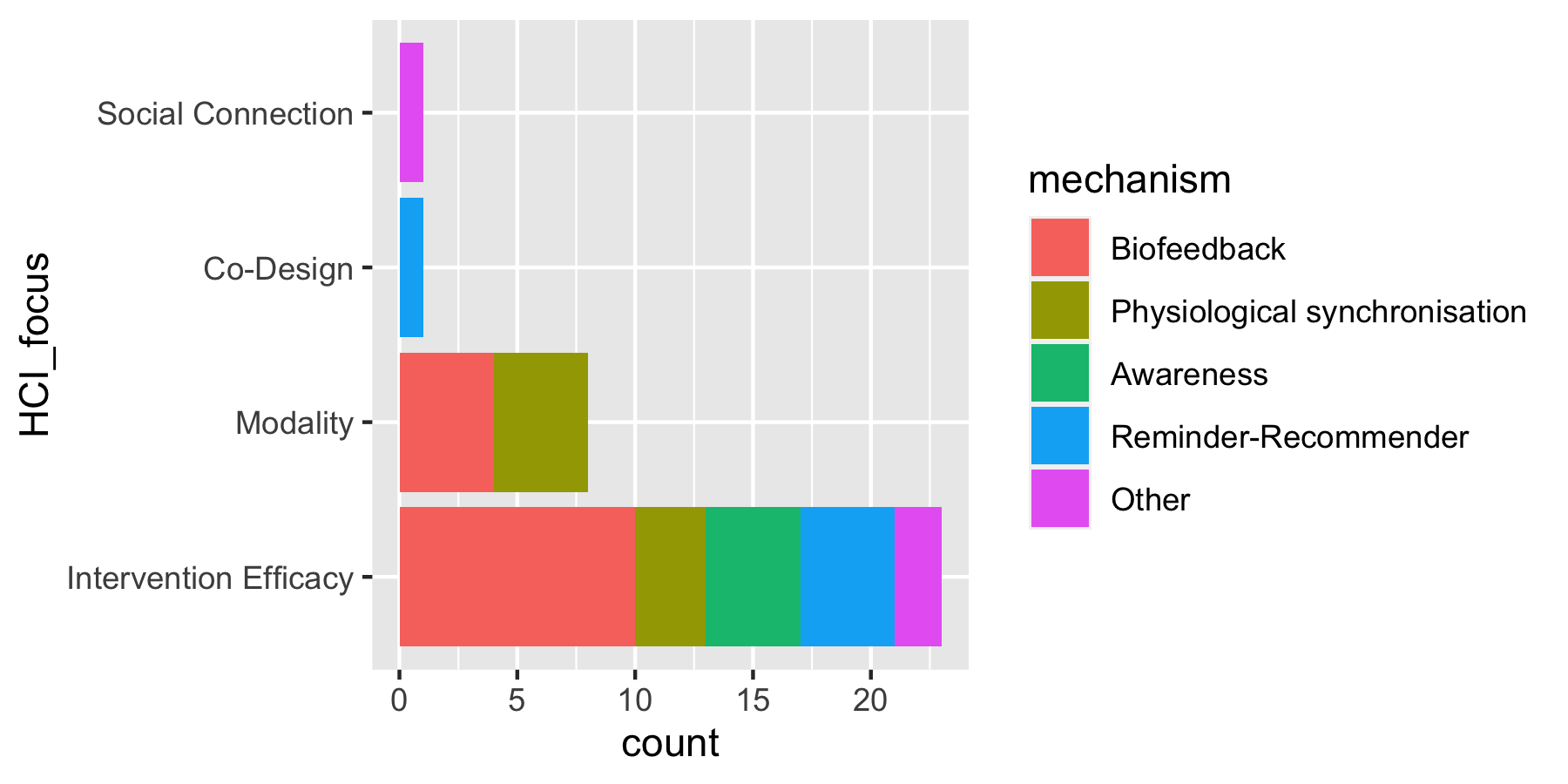}
        \caption{HCI focus by mechanisms}
    \end{subfigure}
    \begin{subfigure}[b]{0.45 \textwidth}
        \centering 
        \includegraphics[width = \textwidth]{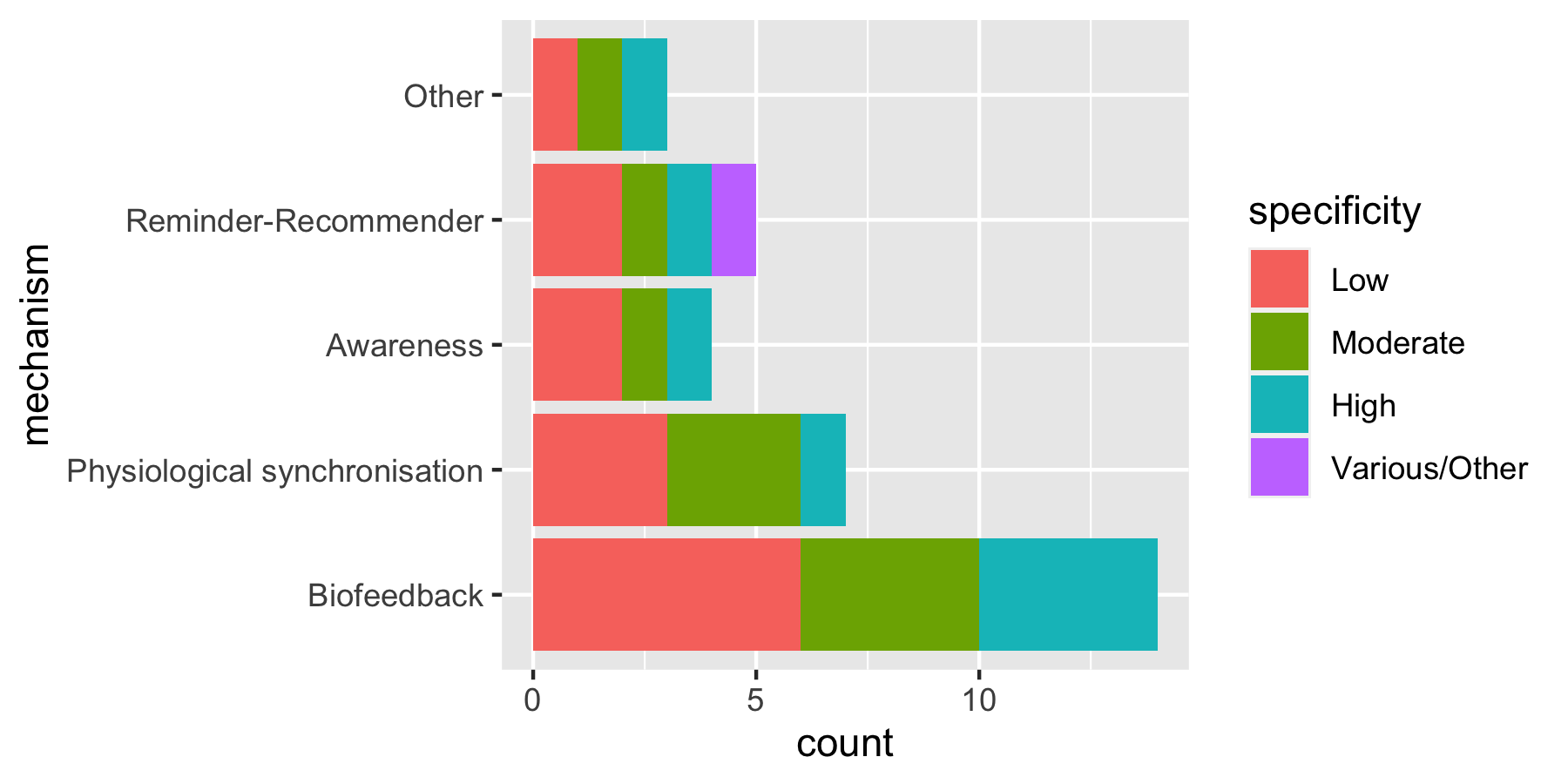}
        \caption{Mechanisms by specificity}
    \end{subfigure}
    \caption{Overview of the interaction between intervention mechanisms, specificity, and HCI focus codes.}
    \label{fig:mechanisms-distribution}
\end{figure}

\subsubsection{Intervention mechanisms} 
\label{sec:appendix1-intervention-mechanisms}
Complementary to the HCI research focus categorisation, the coding of the intervention mechanisms aims to identify the commonly occurring \textit{psychological} mechanisms that the technology-enabled ER intervention work relied on. The purpose is to highlight the importance of the theoretical grounding used---or not used---as part of the intervention design interventions, as well as identify the commonalities and differences across the varied interventions in our dataset.  
We describe the key ideas as well as exemplar papers for each of the category below. In addition, Figure~\ref{fig:mechanisms-distribution} illustrates how the mechanisms and HCI foci intersect. 
\begin{itemize}[label=-, leftmargin = 8pt, itemsep = 4pt, topsep= 4pt, parsep=4pt]

	\item \textit{Biofeedback (n=14)} Biofeedback was the most commonly used intervention mechanism, explored in fourteen papers. Systems created with this model all assume that users will try to consciously or unconsciously modify their behavior as a means to alter their physiological or neurological state closer to a target state. Papers vary in terms of input data source, which may be sensed physiologically (e.g., HVR \cite{Yu2018}) or neurologically (e.g., EEG \cite{Mandryk2013, Antle2018, Antle2019}). Several papers compare input Modalities related to biofeedback design (e.g. \cite{Parnandi2018, Parnandi2017}), and others explore output representations (e.g. light \cite{Yu2018}). Most biofeedback systems are PC-based, however three systems were designed for mobile use \cite{Antle2018, Antle2019, Zafar2017}. There is a group of papers in which biofeedback is embedded into video games as a means to practice ER in the context of game-play (e.g. \cite{Wang2018, Scholten2016, Mandryk2013, Lobel2016}), when these interventions are evaluated for efficacy, research design choices (e.g. lack of control group, no direct measures of ER) often limits validity and/or generalizability of results (see above). Some interventions explicitly provided support to end-users to learn how to modify regulatory responses (e.g. \cite{Knox2011, Antle2018, Antle2019}, but most relied on user’s ability to interpret visual forms of feedback and use that to regulate mental or emotional states. For example, in biofeedback video games, bio-data impacts game mechanics that are communicated to the player through dynamic visual representations (e.g. \cite{Mandryk2013, Scholten2016, Wang2018}). Other biofeedback systems use novel light-based outputs (e.g. \cite{Yu2018} or  visual representations based on metaphor theory (e.g. \cite{Antle2018, Antle2019}) that  provide metaphor-based cues to how to modify mental and/or emotional states. Over time the goal of biofeedback is to learn and practice emotion-regulation, with and without the support of a system until emotion-regulatory behavior becomes increasingly more implicit or automated. Some authors describe explicitly how this transfer might occur (e.g. \cite{Antle2018, Antle2019, Parnandi2018}), while most others do not. 

Few papers explicitly describe how the intervention may result in improved ER skills that last over time. For those that did, biofeedback was often used in combination with or embedded into other frameworks. For example, interventions included biofeedback and psycho-education were proposed as a means to improve learning to modify emotional state and transfer those skills to everyday life (e.g. \cite{Antle2018, Antle2019, Knox2011}). 

    \item \textit{Physiological synchronization (n=7)} papers were based on a system designed based on the mechanism of automatic and implicit emotion regulation that may occur when the individual perceives an external input that mimics their own biosignals (e.g., the `false heart-rate feedback' studies emerging since the 60s \cite{Valins1966}). 
    Systems in this category are designed to digitally represent a target state (e.g. heart rate, breath rate), which is communicated through different modalities to an individual, with the expectation that they will automatically and implicitly synchronized with that target rate. For example, many papers explore the use of haptic representations to prompt a target breath or heart rate (e.g. while driving \cite{Balters2020, Paredes2018}, through a smartwatch \cite{Costa2019, Choi2020} or wearable custom device \cite{Miri2020}).   Four papers explore modalities in early stage research. The others present later stage studies of intervention efficacy, and all have mixed results. Overall, this research area is still in early stages in terms of design choices that might effectively duplicate human-human physiological synchronization as a means to support ER. And, as Choi and Ishii point out, the success of this approach may depends on the individual’s level of sensitivity of interoceptive awareness of their own internal body states related to motion and stress \cite{Choi2020}. In addition, although this approach could result in automatic down-regulation of stress through heart or breath rates, the approach does not posit a mechanism for the explicit development of competency around emotion regulation over time, which could be sustained without the system needing to remain active.
	
	\item \textit{Awareness (n=5)} papers involved monitoring and tracking individuals’ data as a means to increase user’s awareness of their emotional state(s). Input data was gathered from contextual data (e.g., location \cite{Huang2015}) or from self-report, which was entered manually (e.g., diary \cite{Bakker2018}) and/or from sensors (e.g. smartwatch \cite{Wang2018}, multi-sensor wristband \cite{Kocielnik2013}). Displays of information take a variety of forms, largely visual. An underlying assumption in this model is that enhanced awareness will motivate users to take a proactive role in their own stress management and/or that users will reflect on the information being provided to them and as a result of that reflection change their future behaviours. However, none of these papers explore the cause and effect relationship between being shown information and actually implementing strategies for enhanced emotion regulation. In addition, none of the  papers posit a mechanism for the explicit development of competency around emotion regulation, although it is possible that over time an individual might develop competency through repetitive practice.

	\item \textit{Reminders and/or Recommendation (n=6)} papers describe interventions that involve reminders or recommendations to follow step-by-step instructions about how to enact emotion regulation strategies. The papers vary in terms of the complexity of emotion regulation strategies suggested. All systems highlight the need for customized reminders and recommendations. These may be triggered based on manually entered self-reports \cite{Smyth2016,Fage2019,Speer2021} or sensor-based data indicative of stress \cite{Pina2014, Paredes2014} and/or through machine-learning application designed to look for patterns over time \cite{Paredes2014}. Some systems involved content that was customized by others (e.g. teachers, coaches) in terms of strategies \cite{Diaz2018} or in terms of format (e.g. parents) \cite{Fage2019}.  Other content was more generic, for example content was based on strategies from CBT matched to current emotional state \cite{Fage2019}. In one paper the researchers used co-design to explore different forms of recommendations, including the use of physiological data for push \cite{Paredes2014}. Fage et al. \cite{Fage2019} also used participatory design as part of their design process, although it was not the focus of the research. All papers involved evaluations, two at an early formative stage \cite{Pina2014, Diaz2018}, one during deployment during the co-design process \cite{Paredes2014}, and three in more formal deployment \cite{Smyth2016, Fage2019, Speer2021}. One key finding was that for sensor-based push reminders and recommendations the timing rarely coincided with the stressful event due to inaccuracies in biodata to represent stress. In addition, Pina et al suggested that in the moment of stress was not the best timing choice for deliver of reminders and recommendations \cite{Pina2014}.  Findings suggest that manually entered data self-reported data combined with customized and just-in-time content had a positive impact over time (e.g. \cite{Fage2019, Smyth2016}). No papers offer explicit mechanisms that might reduce the need for reminders and recommendations over time.

	\item \textit{Other (n=4)} papers included the use of general relaxation exercises \cite{Carlier2019} and experience-based interventions designed to  provide end-users with the experience of emotion regulation prior to learning how to consciously regulate \cite{Slovak2016, Slovak2018}. One paper presented a system that was mechanisms agnostic, instead instantiating a range of mechanisms into the intervention \cite{Paredes2014}. 
	
\end{itemize}

\subsubsection{Specificity (of the intervention mechanism)}
\label{sec:appendix1-specificity}
In contrast to psychology research, where outlining a clear `theory of change' for the intervention is a key part of the peer review process, HCI work often does not explicitly discuss the presumed causal pathways through which intervention effects should emerge. 
To understand how this (lack of) good practice is present in the emotion regulation intervention space, we coded the level of detail or specificity that was provided about the assumed theory of change \cite{OCathain2019,funnell2011purposeful}: i.e., the cause and effect relationship of the mechanism in relation to prototype and/or intervention design. 

We used a tripartite scale of low, moderate and high level of specificity depending on how clearly and explicitly one or more mechanisms leading to improved emotion regulation were described in the paper and how clearly and explicitly those mechanism(s) were linked to intervention elements posited to create effects related to learning, developing or practicing ER skills. For example, a coding of low would result if a theory of behavior change was described, typically in related work, but no explicit mechanisms were described that could be used to design the intervention. A coding of high was used when one or more mechanisms were clearly explained and explicitly linked to design decisions or  elements of the intervention, typically technological elements. A paper was coded as medium when the mechanisms were present but not well-described and/or the link to design elements was weak. 

 Majority of the intervention mechanisms are described by authors in a low (n = 15) or medium (n = 12) level or of detail. For example, in much of the modality-focused research, many authors assume that the right form of input data linked to particular output representations may lead to enhanced awareness of various aspects of emotion regulation, which in turn should improve emotion regulation. In part, this is based on cognitive behavioral therapy in which awareness of emotional state is posited as a key component of learning ER. However, causal mechanisms or direct linkages for these claims are not described, nor are they already established in the HCI—or psychology—literature. Similarly, much of the R\&R rationale involves the assumption that giving people the right advice at the right time will improve their ability to ER. While physiological synchronization may lead to ER, there is no mechanism posited that relates this to the development of self-regulation skills. Surprisingly, many of the biofeedback intervention mechanisms are also under-specified in terms of how end-users are supposed to learn/know how to change their physiological or neurological states. 

Several successful interventions include multiple mechanisms that together are posited to result in both enacting and developing ER skills that transfer outside of the intervention. For example, \cite{Antle2018,Antle2019,Knox2011} describe one or more intervention mechanisms in a high level of detail, offering  direct explanations for causes and effects in emotion regulation skills development, and provide explicit links from these mechanisms to design features and elements of the intervention, which are subsequently evaluated for efficacy.

\subsubsection{Ongoing support vs Skills development}: 
\label{sec:appendix1-ongoing}
A surprisingly strong majority of the papers in the current data set (n = 31) assumes that the developed systems have to provide an on-going support for the intervention to be effective; and thus the effects would disappear if the system (such as a wearable feedback) was removed. In contrast, only several (n = 5)  interventions were designed to include a way to scaffold or reduce end-user dependence on the intervention and/or specify an explicit transfer mechanism that would enable the end-user to apply emotion regulation strategies outside of the intervention. In \cite{Zafar2017} and \cite{Scholten2016}, the authors proposed that using biofeedback games to teach ER through breath regulation in stressful scenarios may promote carryover to other situations. In \cite{Antle2018} and \cite{Antle2019} the intervention enabled children to directly experience ER through metaphor-based biofeedback in the context of coached interventions that used a CBT approach to promote emotion-regulation skill development and transfer. \cite{Slovak2016} proposed using a shared parent-child experience around an interactive media,  which included narrative hooks to ER strategies taught in school, with the aim to provide parents with language around ER that could be referred to in everyday moments.

\vfill

\vfill~ \pagebreak

\section*{Appendix 2 -- How are the identified design components combined in existing interventions?} 
\label{appendix:clusters}


\begin{figure}[t]
    \centering
    \includegraphics[width=0.9\textwidth]{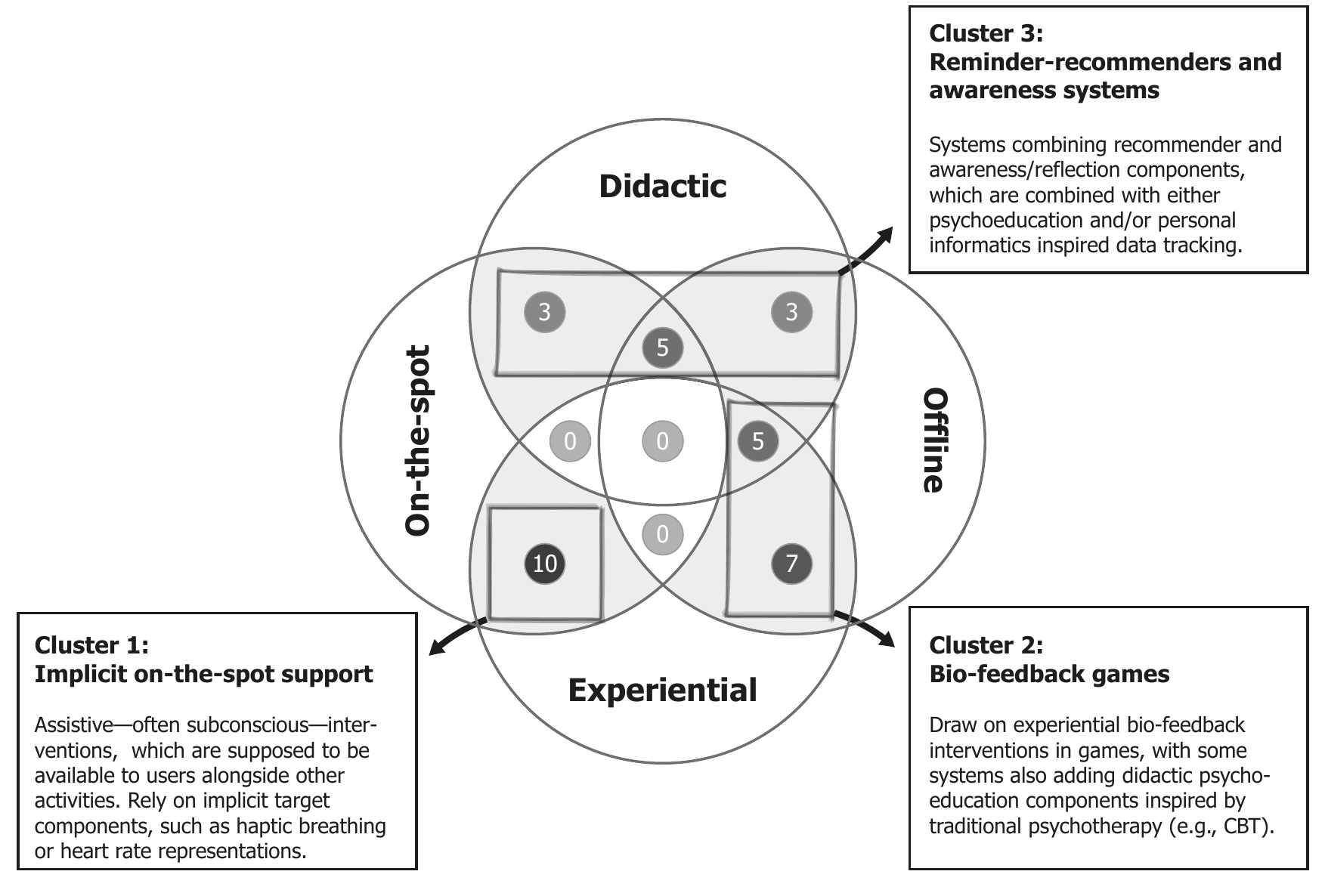} 
    \caption{Illustration of how the reviewed interventions are associated with the four delivery mechanisms -- didactic vs experiential; and offline vs on-the-spot. Numbers and location indicate the number of system using components within the respective intersection of dimensions (e.g., a system using components coded as didactic, on-the-spot, and offline would be placed in the appropriate intersection of the Venn diagram). The annotation shows the resulting clustering of research areas, as described below.}
    \label{fig:dimensionsDiagram}
\end{figure}

The Section \ref{sec:components} synthesised the \textit{design components} present in HCI work so far, with the aim to decompose existing work and provide us with the 'dictionary' of techniques that have been utilised for each of the four possible delivery mechanisms (didactic, experiential, offline, on-the-spot).   
This appendix brings the focus back to the full interventions as presented in prior work to provide an overview of \textit{how the identified components are combined into systems} and identify any meaningful \textit{patterns or gaps of such combinations} across the current dataset. 

The key arguments that we will be making in the rest of the section are as follows:
\begin{enumerate}
\item The existing work is, so far, clearly separated into three main research areas (implicit on-the-spot support, bio-feedback games, and reminders \& awareness systems). These areas are then mostly independent of each other -- they draw on different background literature, intervention goals, and technological components.
\item As a result, the clusters also correspond surprisingly neatly to the combination of delivery mechanisms used, as well as consistent combinations of design components within each cluster (i.e., components used in one cluster---e.g., implicit feedback or recommenders---are rarely utilised in other clusters). 
\item The lack of cross-cluster combinations suggest clear opportunities for future work as described in detail in the next section: for example, there are several combinations of types of intervention components that have not been explored at all so far; as well as more intricate combinations of components that already exist. 
\end{enumerate}

In what follows, we briefly outline the existing research clusters we identified (see Figure~\ref{fig:dimensionsDiagram}) in turn, as well as discuss the currently under-explored segments. 
\bigskip

\subsubsection{Cluster 1 -- implicit on-the-spot support (experiential + on-the-spot) -- n = 11} 

The papers in Cluster 1 comprise systems that explore assistive---often subconscious---down regulation interventions,  which are supposed to be available to users alongside other activities. 
Systems in this category predominantly rely on experiential \textit{implicit target components} (n=8), drawing on a digital representation of a target state (heart rate, breath rate). For example, many papers explore the use of haptic representations of breathing patterns or heart rate (e.g. embedded in driver's seat \cite{Balters2020, Paredes2018}, through a smartwatch \cite{Costa2019, Choi2020} or wearable custom device \cite{Miri2020}).  Other systems use audio to communicate a target breath rate (e.g., \cite{Balters2020}) or a combination of audio and visual representations on a desktop \cite{Ghandeharioun2017}).  Regardless of the target state, \textit{ongoing support} was the most common on-the-spot component: breathing rate was associated with the use of approaches that guide the users toward an appropriate breathing rate through haptic sensations (\cite{Miri2020} or visual cues (\cite{Ghandeharioun2017}), while heartrate feedback was associated with approaches that tap into implicit emotion regulation approaches. Most of this work relies on \textit{situation-specific} intervention delivery, such as in the context of information work (\cite{Moraveji2011,Yu2018}), or driving (\cite{Paredes2018, Balters2020}), and targets nearly exclusively the \textit{response modulation} of the Gross' Process model. 

Three papers showcase the potential for using alternative components to deliver experiential/on-the-spot interventions, predominantly by moving toward more consciously enacted emotion regulation strategies: 
\cite{Yu2018} is aligned with the other work by also providing support alongside other tasks, but relies on \textit{bio-feedback} visualisation to inform participants about their ongoing internal state during a specific task, without any other support for emotion regulation (i.e, targets \textit{identification} rather than directly \textit{response modulation}). 
In contrast, \cite{Liang2018} and \cite{Slovak2018} developed physical objects that the participants can choose to access when needed (\textit{user-initiated} rather than situation-specific delivery). In particular, Liang focused on respiration training through a fidget object including biosensing (a combination of \textit{response modulation} and \textit{emotion awareness}) and Slovak's et al interactive toy then draws on a range of approaches (combining \textit{attention deployment} with in-situ \textit{response modulation} and potential for \textit{cognitive change}). 

\subsubsection{Cluster 2 -- bio-feedback in interactive games (experiential + offline +- didactic) -- n = 12}

The papers in this cluster predominantly draw on utilising experiential bio-feedback interventions in games (n=10), with some systems \cite{Antle2018, Antle2019, Knox2011, Scholten2016} also adding didactic psychoeducation components inspired or directly drawing on traditional psychotherapy (e.g, CBT in-person session with a therapist combined with a bio-feedback game \cite{Knox2011}). 

In terms of the components used, most of the papers rely on the offline components that \textit{both elicit emotion and scaffold the associated emotion regulation} \cite{Lobel2016, Mandryk2013, Parnandi2017, Parnandi2018, Wang2018, Zafar2017, Scholten2016}, utilising a combination of interactive games (used to elicit mostly negative emotions such as stress or fear) and an experiential \textit{real-time biofeedback} loop component to provide feedback scaffolding ER training within the game space. Two systems \cite{Scholten2016, Lobel2016} developed bespoke games including therapeutic techniques into the gameplay. However, the remaining system focused on `only' adding biofeedback as an overlay onto existing games (\cite{Mandryk2013, Wang2018, Parnandi2018, Parnandi2017, Zafar2017}). For example, bio-data was designed to impact game mechanics that are communicated to the player through dynamic visual representations (e.g. \cite{Mandryk2013, Scholten2016, Wang2018}). 
The remaining systems drew on some version of \textit{relaxation training} components \cite{Antle2018, Antle2019, Knox2011}, where the systems utilised traditional biofeedback loop approaches with simple game-like interactions to facilitate users' down-regulation together with formal \textit{psychoeducation} intervention. These games relied on visual representations of the sensed physiological states, using metaphors to visualise the sensed states (e.g., an increase in windy weather) as cues to help modify mental and/or emotional states. Note that a key difference to the elicit\&scaffold model is in the lack of specific emotion generation aspect: the aim is to achieve relaxation (from a baseline state, potentially employing strategies from psycho-education sessions), rather than down-regulate strong emotional experience elicited by the intervention itself. 

In terms of mapping onto the intervention targets from the Process Model of ER, the majority of the game-based interventions are predominantly focused on \textit{response modulation} techniques, with reduction of breathing rate being the most commonly trained emotion regulation technique. In addition, the interventions that include formal psychoeducation components are predominantly CBT based, and thus refer also to \textit{cognitive change} techniques such as cognitive reappraisal. Most of the interventions also do not provide direct support to transfer the learning from the game (offline contexts) into the participants' everyday environment (on-the-spot).  

We note that two papers were exceptions to this overall trend, mostly as they did not describe fully developed interventions:  \cite{Wells2012} is an experimental study comparing the effects of abdominal breathing training (i.e., psychoeducation combined with bio-feedback or no support) on musicians' anxiety in an adapted Trier Social Test; and \cite{Carlier2019}, which is a design exploration of using a series of simple relaxation games (without bio-feedback) to support children with autism (qualitatively tested with 3 participants). 

\subsubsection{Cluster 3 -- recommender \& awareness systems (didactic + offline || on-the-spot) -- n = 13}
The work in this cluster combines the existing work on \textit{recommender} and \textit{awareness and reflection} systems, potentially combined with either \textit{psycho-education} or \textit{visualising patterns over time}.
The papers are more diverse in terms of components used and their combinations in contrast to Clusters 1 \& 2, but this also comes with a less well established groundwork on the assumed theories of change: for example, only 2 out of the 12 papers \cite{Huang2015, Smyth2016} were coded as High on intervention model specificity -- cf., Appendix 1. 

Conceptually, the papers in this cluster can be divided into two main groups, depending on the primary didactic component they rely on (aware\&reflect vs recommender), with two systems having combined both these components \cite{Huang2015,Smyth2016}. 
First, the systems including \textit{aware\&reflect} \cite{Bakker2018, Huang2015, Kocielnik2013, Smyth2016, Wang2019} involved monitoring and tracking individuals’ emotions over time as a means to increase user’s awareness of their emotional state(s). Input data was gathered mostly by participants' EMA reports, which were either self-initiated ('\textit{user-initiated}' on-the-spot component \cite{Huang2015, Wang2019}) or prompted by an automated system ('\textit{system-initiated}' component, \cite{Smyth2016,Bakker2018}); and \cite{Kocielnik2013} relied on an ongoing collection through a wearable sensor. These on-the-spot data collection components were then mostly paired with a reflection interface (offline \textit{visualisation of patterns over time}) to enable the participants to gain new insights based on patterns emerging from the emotional data aggregated over time \cite{Bakker2018,Huang2015,Kocielnik2013,Wang2019}. As such, most of this work is akin to---and likely inspired by---HCI work on personal informatics systems (cf., \cite{Epstein2015}). 

In contrast, the systems relying on \textit{recommenders} provided the users with on-the-spot \textit{suggestions} of specific ER strategies to use \cite{Diaz2018,Fage2019,Huang2015,Paredes2014,Pina2014,Smyth2016, Speer2021}. The content of these recommendations either referred back to what the participants learned in offline \textit{psycho-education} components (mostly traditional talking therapies \cite{Pina2014, Fage2019, Smyth2016}), or `bite-sized' suggestions that were delivered directly as part of the reminder (e.g., activity suggestions in \cite{Diaz2018,Paredes2014,Speer2021}). In most of these systems the specific reminders were contextualised based on the participant's EMA answer \cite{Huang2015,Paredes2014,Smyth2016, Speer2021} or physiological sensing \cite{Pina2014}. Conceptually, these systems often draw on behavioural change intervention systems, especially the Just-in-time-adaptive-intervention literature (see e.g., \cite{Nahum-Shani2018} for a recent review).

In terms of mapping intervention systems onto the intervention targets within the Process Model of ER, the \textit{aware\&reflect} systems are primarily targeting a combination of \textit{self-awareness} and potentially \textit{situation selection} components: e.g., noticing a pattern of negative emotion in a particular situation is assumed to lead to the participant being less likely to engage with similar situations in future iterations.
The \textit{recommender} systems show a broader range of potential intervention targets that are recommended and many refer to content from traditional talking therapies: from \textit{situation selection} (e.g., "go for a run!") to \textit{cognitive change} (e.g., "try to reappraise your emotions") to \textit{response modulation} (e.g., "watch a funny video"). However, this breadth also means that many potentially useful strategies are covered without sufficient depth, especially if the on-the-spot recommendations do not rely on prior psychoeducation components. 


\end{document}